\documentclass[journal]{IEEEtran}

\normalsize

\usepackage{cite}
\usepackage[dvips]{graphicx}
\ifCLASSOPTIONcompsoc
\usepackage[caption=false,font=normalsize,labelfon
t=sf,textfont=sf]{subfig}
\else
\usepackage[caption=false,font=footnotesize]{subfig}
\fi

\usepackage{multicol}
\usepackage[cmex10]{amsmath}
\usepackage{amsthm,amssymb}
\usepackage{bigints}
\usepackage{multirow}
\usepackage{xcolor}
\usepackage{array}
\usepackage{scalerel}
\usepackage{mathtools}

\newtheorem{theorem}{Theorem}
\newtheorem{lemma}{Lemma}

\usepackage{accents}

\begin{document}

\title{Opportunistic Wiretapping/Jamming: A New Attack Model in Millimeter-Wave Wireless Networks}
\author{Yuanyu Zhang, 
		Zhumeng Zheng,
		Ji He, 
		Shuangrui Zhao, 
		Qianyue Qu,
		Yulong Shen,  \IEEEmembership{Member, IEEE},
		and Xiaohong Jiang, \IEEEmembership{Senior Member, IEEE}
\thanks{Y. Zhang, J. He, S. Zhao and Y. Shen are with the School of Computer Science and Technology, Xidian University, Xi'an, Shaanxi, China and also with Shannxi Key Laboratory of Network and System Security. Y. Zhang is also the Beijing Sunwise Information Technology Ltd, Beijing Institute of Control Engineering, Beijing, China. Email: \{yyuzhang, jihe, zhaoshuangrui\}@xidian.edu.cn, ylshen@mail.xidian.edu.cn.}
\thanks{Z. Zheng is with the Beijing Sunwise Information Technology Ltd, Beijing Institute of Control Engineering, Beijing, China. Email: zhengzhumeng@sunwiseinfo.com.}
\thanks{Q. Qu is with the Division of Information Science, Graduate School of Science and Technology, Nara Institute of Science and Technology, Ikoma, Nara, Japan, Email: qu.qianyue.ql2@is.naist.jp.}
\thanks{X. Jiang is with the School of Systems Information Science, Future University Hakodate, Hakodate, Hokkaido, Japan, and also with the School of Computer Science and Technology, Xidian University, Xi'an, Shaanxi, China. Email: jiang@fun.ac.jp.}
}


\makeatletter
\def\ps@IEEEtitlepagestyle{%
  \def\@oddfoot{\mycopyrightnotice}%
  \def\@oddhead{\hbox{}\@IEEEheaderstyle\leftmark\hfil\thepage}\relax
  \def\@evenhead{\@IEEEheaderstyle\thepage\hfil\leftmark\hbox{}}\relax
  \def\@evenfoot{}%
}
\def\mycopyrightnotice{%
  \begin{minipage}{\textwidth}
  \centering \scriptsize
  This work has been submitted to the IEEE for possible publication.  Copyright may be transferred without notice, after which this version may no longer be accessible.
  \end{minipage}
}
\makeatother

\maketitle


\begin{abstract}
While the millimeter-wave (mmWave) communication is robust against the conventional wiretapping attack due to its short transmission range and directivity, this paper proposes a new opportunistic wiretapping and jamming (OWJ) attack model in mmWave wireless networks. With OWJ, an eavesdropper can opportunistically conduct wiretapping or jamming to initiate a more hazardous attack based on the instantaneous costs of wiretapping and jamming. We also provide three realizations of the OWJ attack, which are mainly determined by the cost models relevant to distance, path loss and received power, respectively. To understand the impact of the new attack on mmWave network security, we first develop novel approximation techniques to characterize the irregular distributions of wiretappers, jammers and interferers under three OWJ realizations. With the help of the results of node distributions, we then derive analytical expressions for the secrecy transmission capacity to depict the network security performance under OWJ. Finally, we provide extensive numerical results to illustrate the effect of OWJ and to demonstrate that the new attack can more significantly degrade the network security performance than the pure wiretapping or jamming attack.
\end{abstract}


\IEEEpeerreviewmaketitle

\section {Introduction}\label{sec:introduction} 
\IEEEPARstart{T}{he} past decades have witnessed the explosive growth of wireless devices and the demand for high-speed data traffic, posing a significant challenge to the capacity of wireless communication systems. 
To address this challenge, both industry and academia have advocated communications over the millimeter-wave (mmWave) bands between $30$ and $300$ GHz, where the available bandwidths are orders of magnitude greater than conventional sub-6 bands \cite{Rappaport2013IEEEAccess}. 
In addition, the small wavelength of mmWave signals enables large antenna arrays to be deployed in areas as small as a cellphone or even a chip, achieving significantly high antenna gain for both the transmitting and receiving ends \cite{RanganS2014IEEEProceedings}. Thanks to the above benefits, mmWave communication has been regarded as one of the key enabling technologies in future wireless systems like 5G/6G cellular networks \cite{Hong2021JMicro}, unmanned aerial vehicle (UAV) communications  \cite{ZXiao2022IEEECST} and satellite communications \cite{ZGuo2021IEEETAP}.

Despite the great potential of mmWave communication, its security issue still has been a critical concern due to the broadcast nature of the wireless medium. Physical layer security techniques have been considered as an important approach to ensuring secure mmWave communication \cite{Zhang2019TCOM, SZhao2021TIFS, YXu2021TWC, SZhao2020TIFS, JHe2020TIFS}, since they reveal the fundamental capability of achieving information-theoretic and quantifiable security at the physical layer regardless of the computation capabilities of eavesdroppers. Current research on the physical layer security of mmWave communication mainly focuses on combating the conventional wiretapping attack in various mmWave systems, such as mmWave multiple-input and multiple-output (MIMO) systems \cite{ZKong2021TIFS, JXU2019TCOM, KWu2020ComLett, WHao2020TCOM, JSong2020TvT}, mmWave non-orthogonal multiple access (NOMA) systems  \cite{SHuang2020TvT, YSong2020TvT, YYapici2021TvT, XSun2020IoTJ}, intelligent reflecting surface (IRS)-aided mmWave systems \cite{YZhu2020JSAC, CWang2021TvT, YXiu2021ComLett}, hybrid mmWave-free space optical (FSO) systems  \cite{Shakir2022PhotonicsJ, Tokgoz2022TvT, Tokgoz2021ICC, Saber2021SystJ} and mmWave ad hoc networks \cite{YZhang2022TIFS, YZhu2017TWC, YZhu2018JSAC}, etc (See Section \ref{sec:related-work} for related works).

The aforementioned works help us to understand the impacts of the conventional wiretapping attack on the security performance of mmWave wireless systems, while the recent real-world measurements on mmWave propagation characteristics indicate that mmWave signals are actually less susceptible to the conventional wiretapping attack \cite{Rappaport2017TAP, ma2018security, akyildiz2014terahertz,harvey2019exploiting}. This is mainly due to the following two reasons. First, mmWave signals suffer from severe atmospheric absorption, rain attenuation and penetration loss, leading to a limited transmission range and also the intermittent connectivity of wiretapping links (especially when the links are blocked by obstacles like buildings and human bodies). Second, to combat the severe signal attenuation, transmitters are usually equipped with highly directional antennas with narrow beams, rendering wiretapping on mmWave links much more difficult.

To further identify the potential threats to mmWave communication security, this paper integrates the jamming technique with wiretapping and proposes a more hazardous opportunistic wiretapping and jamming (OWJ) attack model in a mmWave ad hoc network consisting of multiple transmitters, receivers and eavesdroppers distributed according to Poisson Point Processes (PPPs). With OWJ, an eavesdropper can opportunistically conduct wiretapping or jamming based on the instantaneous costs of conducting wiretapping and jamming, aiming at achieving a more significant attack effect. This paper extends its conference version in \cite{YZhang2021NaNA} by adding more OWJ attack realizations and secrecy performance analysis. The main contributions of this paper are summarized as follows.

\begin{itemize}
\item By combining the jamming technique with wiretapping, we propose a new and more hazardous OWJ attack model in mmWave wireless networks, which allows eavesdroppers to conduct wiretapping and jamming opportunistically. This model covers the conventional wiretapping attack and jamming attack as special cases. We also provide three realizations of the OWJ attack, namely distance-related OWJ (DOWJ), loss-related OWJ (LOWJ) and power-related OWJ (POWJ), where the cost of an eavesdropper in wiretapping/jamming is characterized by the distance, path losses and received power, respectively.
\item 
To understand the impact of the OWJ attack on mmWave network security, we first develop novel approximation techniques to characterize the irregular distributions of the wiretappers around a target transmitter, the jammers around a target legitimate receiver and the interferers around a target wiretapper under the three OWJ realizations. With the help of the results of these node distributions, we then derive analytical expressions for the secrecy transmission capacity (STC) to depict the network security performance under the OWJ attack model.
\item We finally provide numerical results to illustrate the network STC performance under the OWJ attack model. The results revealed that, in general, the OWJ can more significantly degrade the network security performance than the pure jamming or pure wiretapping attack. In particular, among the three realizations, the POWJ serves as the most hazardous attack, while the DOWJ and LOWJ lead to almost the same attack effect.
\end{itemize}

The remainder of this paper is organized as follows. We present the related work regarding security performance analysis in mmWave systems under the conventional wiretapping attack in Section \ref{sec:related-work}. We introduce the system model in Section \ref{sec:sys-model} and conduct theoretical performance analysis in Sections \ref{sec:cp-analysis} and \ref{sec:sp-analysis}.
Section \ref{sec:num-res} presents the numerical results and discussions, and Section \ref{sec:con} concludes this paper.

\section{Related Work}\label{sec:related-work}
Existing works on combating the conventional wiretapping attack in mmWave systems mainly focus on the corresponding system security evaluation therein. 
We categorize these works according to the considered system models as follows.
\subsection{MmWave MIMO Systems}
MIMO has been regarded as an appealing technology to reap the benefits of mmWave communication, motivating a plethora of work investigating the security performance of mmWave MIMO systems \cite{ZKong2021TIFS, JXU2019TCOM, KWu2020ComLett, WHao2020TCOM, JSong2020TvT}.
For instance, the joint design of the analog and digital precoders of the secondary transmitter (ST) was investigated to maximize the secrecy rates of the secondary users (SRs) in a mmWave cognitive network with one ST broadcasting information to multiple SRs in the presence of multiple primary users and eavesdroppers \cite{ZKong2021TIFS}.
A sparse mmWave massive MIMO network was considered in \cite{JXU2019TCOM}, where a security scheme was proposed to send information signals through dominant angles of the sparse channel and broadcast AN over the remaining non-dominant angles to interfere only with the eavesdropper. The optimal sparsity parameter was determined to maximize the network secrecy rate.
In \cite{KWu2020ComLett}, the secrecy rate was analyzed for a mmWave lens antenna array transmission system and the optimal power allocation between signal and AN was explored to maximize the system secrecy.
The optimal beamformer design was investigated for various scenarios, like the downlink communication of mmWave cloud radio access networks (C-RANs) \cite{WHao2020TCOM} and the dual-polarized mmWave communication systems \cite{JSong2020TvT}.

\subsection{MmWave NOMA Systems} 
The combination of the NOMA technology and mmWave communication for enhanced security has recently been a hot research topic \cite{SHuang2020TvT, YSong2020TvT, YYapici2021TvT, XSun2020IoTJ}.
The authors in \cite{SHuang2020TvT} considered the downlink transmission in a mmWave NOMA network and proposed a minimal angle-difference user paring scheme and two maximum ratio transmission beamforming schemes to enhance the network security.
In \cite{YSong2020TvT}, a cognitive mmWave NOMA network was introduced, where each resource block is shared by a user pair consisting of a primary user and a secondary user. The security and reliability performances of the secondary users were investigated under a power allocation scheme that prioritizes the security and QoS requirements of the primary users.
The authors in \cite{YYapici2021TvT} focused on a UAV-to-ground communication scenario, where the UAV base station (BS) adopts the NOMA technology to serve multiple ground users via mmWave downlinks. They proposed a protected-zone approach to suppress the eavesdroppers outside the user region and analyzed the system security performance in terms of the secrecy rate.
The NOMA-assisted mmWave UAV downlink scenario was also considered in \cite{XSun2020IoTJ}, whereas, different from  \cite{YYapici2021TvT}, part of the ground users are energy-constrained and thus use a portion of the power from the BS for energy harvesting and the remaining power for information processing. 

\subsection{MmWave IRS Systems}
The IRS technology, also known as reconfigurable intelligent surface (RIS), has also been introduced to enhance the security of mmWave communication systems recently \cite{YZhu2020JSAC, CWang2021TvT, YXiu2021ComLett}.
The authors in \cite{YZhu2020JSAC} considered a mmWave network consisting of BSs, users and RISs, where the locations of BSs and RISs are modeled by homogeneous PPPs. 
They proposed a two-step association rule to link BSs, users and RISs, and analyzed the area spectrum efficiency and energy efficiency to show the performance gain achieved by the RISs.
In \cite{CWang2021TvT}, an IRS was introduced to assist the BS in securely broadcasting information to users while sending its own information to an IoT device. 
The precoder at the BS and the beamformer at the IRS were jointly designed to maximize the minimum secrecy rate of multiple users.
In \cite{YXiu2021ComLett}, the secrecy rate maximization problem was solved for a RIS-aided massive MIMO mmWave downlink scenario with constrained hardware cost and power budget at the transmitter.

\subsection{Hybird mmWave-FSO Systems}
As a promising candidate for backhaul solutions of 5G and beyond 5G networks, hybrid communication systems with mmWave links coexisting with FSO links have attracted considerable attention, and so does the security performance therein \cite{Shakir2022PhotonicsJ, Tokgoz2022TvT, Tokgoz2021ICC, Saber2021SystJ}.
The authors in \cite{Shakir2022PhotonicsJ} conducted security performance analysis in terms of average secrecy capacity, secrecy outage probability and strictly positive secrecy capacity in a parallel FSO-mmWave communication system, where the transmitter sends information to the receiver via an FSO link and a mmWave link concurrently. Three wiretapping scenarios were considered, where the FSO link, mmWave link or both links are wiretapped, respectively. 
A similar scenario was considered in \cite{Tokgoz2022TvT} but with different channel fading randomness, i.e., exponential atmospheric turbulence and Weibull fading channels.
The parallel FSO-mmWave system was also considered in \cite{Tokgoz2021ICC}, while, unlike \cite{Shakir2022PhotonicsJ}, the authors focused on a wiretapping channel that is correlated to the main channel and hybrid eavesdroppers that can wiretap both the FSO and mmWave links. 
Apart from the parallel FSO-mmWave system, a serial two-hop FSO-mmWave system consisting of one FSO link and one mmWave link has been introduced in \cite{Saber2021SystJ}, where the average secrecy capacity and secrecy outage probability were analyzed for system security performance analysis.

\subsection{mmWave Ad Hoc Networks}
Research efforts have also been devoted to the security performance analysis of mmWave ad hoc networks. 
The authors in \cite{YZhang2022TIFS} considered a mmWave ad hoc network with transmitters, receivers, potential jammers and eavesdroppers distributed according to PPPs. They proposed a sight-based cooperative jamming scheme, where each potential jammer that has a non-Line-of-Sight (NLoS) link to its nearest receiver but may have LoS links to eavesdroppers is selected with a certain probability to radiate AN such that channel advantages at the receivers can be achieved. The network STC performance was analyzed to show the secrecy gain achieved by the proposed scheme.
In \cite{YZhu2017TWC}, a similar scenario without potential jammers was considered and the average achievable secrecy rate was analyzed under a simple AN-based transmission scheme, where each transmitter allocates a fraction of its transmit power to radiate AN. 
A 3D ad hoc network was considered in \cite{YZhu2018JSAC}, where UAVs transmit to ground receivers in the presence of ground eavesdroppers. 
The secrecy rate performance was investigated under a simple cooperative jamming scheme that allows part of the UAVs to radiate AN without taking their link conditions to the ground receivers into consideration. 

\section{System Model} \label{sec:sys-model}
In this section, we introduce the network, antenna, blockage and propagation models, followed by the proposed OWJ attack model and the performance metrics.
\subsection{Network Model}
We consider a Poisson bipolar network comprising a set of mmWave transmitter-receiver pairs, where each receiver is located at a fixed distance $r_0$ away from its transmitter but at random orientation.
The locations of the transmitters and receivers are modeled by two \emph{dependent} PPPs $\Phi_T$ and $\Phi_R$ of the same intensity  \cite{haenggi2012stochastic}, denoted by $\lambda$. 
To simplify the analysis, we neglect the dependence between $\Phi_T$ and $\Phi_R$, which still achieves accurate approximations, as can be seen from \cite{YZhang2022TIFS} and the results in this paper.
Also present in the network is a set of \emph{half-duplex} eavesdroppers, whose locations are modeled by another independent and homogenous PPP $\Phi_E$ of intensity $\lambda_E$. 
Each eavesdropper independently selects to conduct the \emph{wiretapping} attack or the \emph{jamming} attack based on a certain OWJ strategy, as introduced in Section \ref{sec:sa-strategy}.
The resulting wiretappers and jammers form two independent PPPs, denoted by $\Phi_W$ and $\Phi_J$, respectively. 
Using $\mathbf{1}_J^z$ to indicate whether an eavesdropper $z$ is a jammer (i.e., $\mathbf{1}_J^z=1$) or not (i.e., $\mathbf{1}_J^z=0$), we have
\begin{IEEEeqnarray}{rCl}
\Phi_J=\{z\in \Phi_E:\mathbf{1}_J^z=1\}, \Phi_W=\{z\in \Phi_E:\mathbf{1}_J^z=0\}.
\end{IEEEeqnarray}

\subsection{Antenna Model}\label{sec:antenna-model}
Each node is equipped with an antenna array to form a directional antenna.
To approximate such antennas, we adopt the sectored antenna model \cite{YZhang2022TIFS, Zhang2019TCOM, SZhao2021TIFS, YXu2021TWC, SZhao2020TIFS, JHe2020TIFS}, where each antenna consists of a main lobe and a side lobe.
The key antenna parameters of different node types are summarized in Table \ref{tb:antenna-para}.
\begin{table}[t]
\caption{Antenna Parameters (ML: Main Lobe, SL: Side Lobe)}
\renewcommand{\arraystretch}{1.5}
\begin{center} \label{tb:antenna-para}
\begin{tabular}{ | m{5.5em} | m{1.3cm} | m{1.2cm} | m{1.2cm} | m{1.2cm} | } 
\hline
\bf{Parameters} & \bf{Transmitter} & \bf{Receiver} & \bf{Wiretapper} & \bf{Jammer}\\ 
\hline
ML width & $\theta_T$ & $\theta_R$ & $\theta_W$ & $\theta_J$\\
\hline
SL width & $2\pi-\theta_T$ & $2\pi-\theta_R$ & $2\pi-\theta_W$ & $2\pi-\theta_J$\\
\hline
ML  gain & $G_M^T$ & $G_M^R$ & $G_M^W$ & $G_M^J$\\
\hline
SL gain & $G_S^T$ & $G_S^R$ & $G_S^W$ & $G_S^J$\\
\hline
\end{tabular}
\end{center}
\end{table}
Due to the isotropic feature of the PPPs, the effective antenna gain between a transmitting node $a$ of type $t_1\in \{T,J\}$ ($T$: transmitter, $J$: jammer) and a receiving node $b$ of type $t_2\in\{R,W\}$ ($R$: receiver, $W$: wiretapper) can be represented by the following random variable
\begin{align} \label{eqn:antenna-gain-m}
\mathsf{G}_{t_1,t_2}^{a,b} =
  \begin{cases}
    G_M^{t_1} G_M^{t_2}, &  \text{ w.p. } p_{MM}^{t_1t_2}=\frac{\theta_{t_1}\theta_{t_2}}{(2\pi)^2}\\
    G_M^{t_1} G_S^{t_2},  & \text{ w.p. } p_{MS}^{t_1t_2}=\frac{\theta_{t_1} (2\pi-\theta_{t_2})}{(2\pi)^2} \\
    G_S^{t_1} G_M^{t_2},  & \text{ w.p. } p_{SM}^{t_1t_2}=\frac{(2\pi-\theta_{t_1})\theta_{t_2}}{(2\pi)^2} \\
    G_S^{t_1} G_S^{t_2},   & \text{ w.p. } p_{SS}^{t_1t_2}=\frac{(2\pi-\theta_{t_1})(2\pi-\theta_{t_2})}{(2\pi)^2} 
  \end{cases},
\end{align}
where $ \text{ w.p. }$ stands for \emph{with probability}.
Prior to transmissions, each pair of transmitter and receiver align their antennas to achieve the largest antenna gain $G_M^T G_M^R$.

\subsection{Blockage and Propagation Model}
Due to the existence of blockages, communication links can be LoS or NLoS.
According to the blockage model in \cite{AThornburg2016TSP}, a link of length $r$ is LoS with probability $p_{L}(r)=e^{-\beta r}$ and NLoS with probability $p_{N}(r)=1-p_{L}(r)$, where $\beta$ denotes the blockage density.
We use $\mathcal S_{a,b}$ to represent the status of the link $a\rightarrow b$ between nodes $a$ and $b$.
$\mathcal S_{a,b}=L$ (resp. $\mathcal S_{a,b}=N$) means that the link is LoS (resp. NLoS).
Links suffer from both large-scale path loss and small-scale fading characterized by the Nakagami fading model.
The path loss of the link $a\rightarrow b$ is $d_{a,b}^{\alpha}$, where $d_{a,b}$ denotes the distance between nodes $a$ and $b$, and $\alpha$ is the random path-loss exponent, which equals $\alpha_L$ (resp. $\alpha_N$) if $\mathcal S_{a,b}=L$ (resp. $\mathcal S_{i,j}=N$) . 
The corresponding channel gain $h_{a,b}$ follows the gamma distribution with shape $N$ and rate $N$.
Here, $N=N_L$ (resp.  $N=N_N$) if $\mathcal S_{a,b}=L$ (resp. $\mathcal S_{a,b}=N$).
Throughout this paper, we assume $\alpha_L<\alpha_N$ and $N_L>N_N$.

\subsection{OWJ Attack Molde}\label{sec:sa-strategy}
In the OWJ attack, each eavesdropper measures the costs of wiretapping and jamming and conducts the wiretapping attack if
$$\mathrm{cost~of~wiretapping} < \rho \cdot \mathrm{cost~of~jamming},$$
and conducts the jamming attack otherwise.
The bias factor $\rho$ here represents the preference of the eavesdroppers for the wiretapping attack.
The larger the $\rho$ is, the more likely eavesdroppers will wiretap. 
Note that the OWJ attack covers the pure wiretapping (resp. jamming) attack, as $\rho$ tends to $\infty$ (resp. $0$).
In this paper, we consider the following three representations of the costs of a typical eavesdropper $z\in\Phi_E$, giving rise to three different realizations of the OWJ attack, i.e., DOWJ, LOWJ, POWJ, respectively.
This is motivated by the fact that eavesdroppers manage to improve their attack effect with all the available network knowledge.
\begin{itemize}
\item \emph{Smallest distances}: We use the smallest distances from $z$ to the transmitters and receivers to represent the costs of wiretapping and jamming, which are denoted by $\mathsf{D}_T^z=\min_{x\in\Phi_T}d_{x,z}$ (wiretapping) and $\mathsf{D}_R^z=\min_{y\in\Phi_R}d_{y,z}$ (jamming), respectively. 
This applies to the case where only the location information of the transmission pairs is known to the eavesdroppers. 
\item \emph{Smallest path losses}: We use the smallest path losses from $z$ to the transmitters and receivers as the costs. Formally, the costs are given by $\mathsf{L}_T^z=\min_{x\in\Phi_T}d_{x,z}^{\alpha}$ (wiretapping) and  $\mathsf{L}_R^z=\min_{y\in\Phi_R}d_{y,z}^{\alpha}$ (jamming), respectively.
This applies to the case where both the locations of the transmission pairs and the link status to the transmission pairs are known to the eavesdroppers.
\item \emph{Smallest reciprocals of power}: We use the smallest reciprocal of the power received by $z$ (resp. receivers) from the transmitters (resp. $z$) as the cost of wiretapping (resp. jamming). 
The costs are formally given by $\mathsf{P}_T^z=\min_{x\in\Phi_T}d_{x,z}^{\alpha}/(P_T \mathsf{G}_{T,W}^{x,z})$ (wiretapping) and $\mathsf{P}_R^z=\min_{y\in\Phi_R}d_{y,z}^{\alpha}/(P_J \mathsf{G}_{J,R}^{z,y})$ (jamming), where $P_T$ and $P_J$ denote the transmit power and jamming power of the transmitters and jammers, respectively. 
This applies to the case where the information of instantaneous antenna gains to the transmitters and receivers is also available.
\end{itemize}

The three OWJ attack realizations cover scenarios with different network knowledge available to the eavesdroppers.
By analyzing the security performance under the attacks, we can identify the knowledge that has a significant impact on the attack effect.

\subsection{Performance Metrics}
Transmitters adopt the Wyner encoding scheme \cite{YZhang2022TIFS} for transmissions, where each confidential message is encoded into a codeword that is randomly selected from multiple candidates. 
Such randomness is used to confuse eavesdroppers.
Two code rates are defined in this scheme, i.e., the code rate for the codeword $R_t$ and that for the confidential message $R_s$.
The difference $R_e=R_t-R_s$ reflects the code rate sacrificed for generating the randomness.
We assume that $R_t$, $R_s$ and $R_e$ are fixed throughout this paper.

We adopt the STC metric to model the security performance, which defines the average sum rate of transmissions in perfect secrecy per unit area. Formally, the STC is given by 
\begin{align}\label{eqn:def-stc}
C_{s}=\lambda p_{c}p_{s}R_s,
\end{align}
where $p_{c}$ denotes the connection probability of transmissions (i.e., the probability that receivers can successfully recover the confidential messages), $p_{s}$ denotes the secrecy probability of transmissions (i.e., the probability that the eavesdroppers fail to decode the confidential messages).
We can see from \eqref{eqn:def-stc} that $p_c$ and $p_s$ are the key parameters for determining the STC.
Thus, we focus on the analyses of these two probabilities in the subsequent sections.

\section{Connection Probability Analysis}\label{sec:cp-analysis}
In this section, we first derive a unified expression for the connection probability of a typical transmission pair $x_0\rightarrow y_0$, which involves a key term, i.e., the Laplace transform of the interference at $y_0$ from the jammers.
We then derive this Laplace transform under the DOWJ, LOWJ and POWJ attacks in Sections \ref{sec:ly0j-sad}, \ref{sec:ly0j-sal} and \ref{sec:ly0j-sap}, respectively.

\subsection{Connection Probability}
According to the definition, the connection probability is
\begin{IEEEeqnarray}{rCl}\label{eqn:pc-def}
p_c=\mathbb P\big(\log(1+\mathrm{SINR}_{x_0,y_0})>R_t\big),
\end{IEEEeqnarray}
where $\mathrm{SINR}_{x_0,y_0}$ is the signal-to-interference-plus-noise ratio (SINR) of $y_0$ and given by
\begin{IEEEeqnarray}{rCl}\label{eqn:sinrx0y0}
\mathrm{SINR}_{x_0,y_0}=\frac{P_T G_M^TG_M^R h_{x_0,y_0}r_0^{-\alpha}}{I_{y_0}^{T}+I_{y_0}^J+\sigma^2}.
\end{IEEEeqnarray}
Here, $\sigma^2$ is the noise power at $y_0$ and $I_{y_0}^{T}$ (resp. $I_{y_0}^{J}$) denotes the interference at $y_0$ caused by the transmitters in $\Phi_T$ (resp. jammers in $\Phi_J$).
Based on \eqref{eqn:pc-def} and \eqref{eqn:sinrx0y0}, $p_c$ can be derived as follows.
\begin{theorem}
The unified connection probability of the typical transmission pair $x_0\rightarrow y_0$ under the DOWJ, LOWJ and POWJ attacks can be approximated by 
\begin{IEEEeqnarray}{rCl}\label{eqn:pc}
p_c&\approx&\!\!\!\!\sum_{k\in\{L,N\}}\!\!\!\!\!p_k(r_0)\!\sum_{t=1}^{N_k}\!\binom{N_k}{t}\!(-1)^{t+1}\! e^{-t\mu_k\sigma^2}\!\!\!\!\!\!\prod_{t_1\in \{T,J\}}\!\!\!\!\!\!\mathcal{L}_{y_0}^{t_1}(t\mu_k),
\end{IEEEeqnarray}
where $\mu_k=N_k(N_k!)^{-1/N_k} (2^{R_t}-1)r_0^{\alpha_k}/(P_T G_M^TG_M^R)$ and $\mathcal{L}_{y_0}^{t_1}(\cdot)$ ($t_1\in\{T,J\}$) denotes the Laplace transforms of $I_{y_0}^{t_1}$.
\end{theorem}
\begin{IEEEproof}
Conditioned on the event that the link $x_0\rightarrow y_0$ is in status $k$ (i.e., $\mathcal S_{x_0,y_0}=k, k\in\{L,N\}$), $p_c$ is given by
\begin{IEEEeqnarray}{rCl}
p_c^k&\approx&\sum_{t=1}^{N_k}\binom{N_k}{t}(-1)^{t+1} e^{-t\mu_k\sigma^2}\prod_{t_1}\mathcal{L}_{y_0}^{t_1}(t\mu_k),
\end{IEEEeqnarray}
according to Theorem $1$ in \cite{YZhang2022TIFS}.
Applying the law of total probability in terms of the link status completes the proof.
\end{IEEEproof}
We can see from \eqref{eqn:pc} that the key terms to determine $p_c$ is the Laplace transforms $\mathcal{L}_{y_0}^{t_1}(\cdot)$.
Note that $\mathcal{L}_{y_0}^{T}(\cdot)$ is independent of the OWJ attack model, while $\mathcal{L}_{y_0}^{J}(\cdot)$ is not.
Prior to deriving $\mathcal{L}_{y_0}^{T}(\cdot)$, we present the following functions for any $k\in\{L,N\}$, $i\in\{M,S\}$, $j\in\{M,S\}$, $t_1\in \{T,J\}$ and  $t_2\in\{R,W\}$, which will be used extensively in this paper.
\begin{IEEEeqnarray}{rCl}\label{eqn:omegat1t2kij}
\Omega_{t_1,t_2}^{k,i,j}(s,r)=p_{ij}^{t_1t_2}p_k(r)\!\!\left(\!1\!-\!\left(1\!+\!\frac{sP_{t_1}G_i^{t_1} G_j^{t_2}}{N_k r^{\alpha_k}}\right)^{-N_k}\!\right),
\end{IEEEeqnarray}
\begin{IEEEeqnarray}{rCl}\label{eqn:omegat1t2k}
\Omega_{t_1,t_2}^{k}(s,r)=\sum_{i,j}\Omega_{t_1,t_2}^{k,i,j}(s,r),
\end{IEEEeqnarray}
\begin{IEEEeqnarray}{rCl}\label{eqn:omegat1t2}
\Omega_{t_1,t_2}(s,r)=\sum_{k}\Omega_{t_1,t_2}^{k}(s,r).
\end{IEEEeqnarray}

Now, we are ready to derive $\mathcal{L}_{y_0}^{T}(\cdot)$ in the following lemma.
\begin{lemma}\label{lemma:ly0t}
The Laplace transform of $I_{y_0}^T$ under the DOWJ, LOWJ and POWJ attacks is
\begin{IEEEeqnarray}{rCl}\label{eqn:ly0t}
\mathcal {L}_{y_0}^T(s)&=&\exp\left(-2\pi \lambda \int_{0}^{\infty}\Omega_{T, R}(s,r)r \mathrm dr \right).
\end{IEEEeqnarray}
\end{lemma}
\begin{IEEEproof}
According to the definition, we have
\begin{IEEEeqnarray}{rCl}
\mathcal {L}_{y_0}^T(s)\!&=&\!\mathbb E\left[e^{-sI_{y_0}^T}\right]\nonumber\\
\!&=&\!\mathbb E\left[e^{-s\sum_{x\in\Phi_T\backslash\{x_0\}} P_T\mathsf{G}_{T,R}^{x,y_0} h_{x,y_0}d_{x,y_0}^{-\alpha}}\right]\nonumber\\
\!&=&\!\mathbb E_{\Phi_T}\left[\prod_{x\in\Phi_T\backslash\{x_0\}}\mathbb E_{\mathsf{G}_{T,R}^{x,y_0},h_{x,y_0},\alpha}\left[e^{-\frac{sP_T\mathsf{G}_{T,R}^{x,y_0} h_{x,y_0}}{d_{x,y_0}^{\alpha}}}\right]\right]\nonumber\\
\!&\overset{(a)}{=}&\!\exp\Bigg(\!-\!2\pi \lambda\int_{0}^{\infty}\!\Bigg(1\!-\!\mathbb E\left[e^{-\frac{sP_T\mathsf{G}_{T,R}^{x,y_0} h_{x,y_0}}{r^{\alpha}}}\right]\!\Bigg)r\mathrm d r\!\Bigg)\nonumber\\
&=&\exp\left(-2\pi \lambda \int_{0}^{\infty}\Omega_{T, R}(s,r)r \mathrm dr \right),
\end{IEEEeqnarray}
where $(a)$ follows after applying the probability generating functional of PPP \cite{haenggi2012stochastic}.
\end{IEEEproof}
Next, we derive $\mathcal{L}_{y_0}^{J}(\cdot)$ under the three OWJ attacks in the following subsections, respectively.

\subsection{Derivation of $\mathcal{L}_{y_0}^{J}(s)$ under DOWJ Attack} \label{sec:ly0j-sad}
Before deriving $\mathcal{L}_{y_0}^{J}(s)$, we first establish the following lemma for the probability that an eavesdropper $z$ with distance $v$ to $y_0$ is a jammer (i.e., $\mathbf{1}_J^z=1$) under the DOWJ attack.
\begin{lemma}\label{lemma:zeta-sad}
The probability that an eavesdropper $z$ with distance $v$ to the typical receiver $y_0$ is a jammer under the DOWJ attack is
\begin{IEEEeqnarray}{rCl}\label{eqn:zeta}
\zeta(v)&=& \frac{1}{\rho^2+1}+\frac{\rho^2}{\rho^2+1}e^{-(\rho^2 +1)\lambda\pi v^2}\bar{F}_{d_{x_0,z}}(c_v)\nonumber\\
&&-\int_{\lvert v-r_0\rvert}^{c_v}\frac{e^{-(1 +\frac{1}{\rho^2})\lambda\pi u^2}}{\rho^2+1}f_{d_{x_0,z}}(u)\mathrm du,
\end{IEEEeqnarray}
where $c_v=\min\{\max\{\rho v, \lvert v-r_0\rvert\}, v+r_0\}$, 
\begin{IEEEeqnarray}{rCl}\label{eqn:ccdf_dx0z}
\bar{F}_{d_{x_0,z}}(u)&=&1-\frac{1}{\pi}\arccos\left(\frac{r_0^2+v^2-u^2}{2r_0v}\right)
\end{IEEEeqnarray}
is the complementary cumulative distribution function (CCDF) of $d_{x_0,z}$ and
\begin{IEEEeqnarray}{rCl}\label{eqn:pdf_dx0z}
f_{d_{x_0,z}}(u)&=&\frac{2u}{\pi\sqrt{4r_0^2v^2-(r_0^2+v^2-u^2)^2}}
\end{IEEEeqnarray}
is the corresponding probability density function (PDF).
\end{lemma}
\begin{IEEEproof}
See Appendix \ref{app:proof-lemma-zeta-sad}.
\end{IEEEproof}

Based on Lemma \ref{lemma:zeta-sad}, we derive $\mathcal{L}_{y_0}^{J}(s)$ under the DOWJ attack in the following lemma.
\begin{lemma}\label{lemma:ly0j-sad}
The Laplace transform of $I_{y_0}^J$ under the DOWJ attack can be lower bounded by 
\begin{IEEEeqnarray}{rCl}\label{eqn:iy0j-sad}
\mathcal {L}_{y_0}^J(s)&\ge&\exp\left(-2\pi \lambda_E\int_{0}^{\infty}\Omega_{J,R}(s,v)\zeta(v)v\mathrm d v\right).
\end{IEEEeqnarray}
\end{lemma}
\begin{IEEEproof}
First, we have
\begin{IEEEeqnarray}{rCl}\label{eqn:iy0j}
I_{y_0}^J=\sum_{z\in\Phi_E } \mathbf{1}_J^zP_J \mathsf{G}_{J,R}^{z,y_0} h_{y_0,z}d_{y_0,z}^{-\alpha}.
\end{IEEEeqnarray}
Hence,
\begin{IEEEeqnarray}{rCl}
\mathcal {L}_{y_0}^J(s)\!&=&\!\mathbb E_{I_{y_0}^J}\left[e^{-sI_{y_0}^J}\right]\\
&=&\mathbb E_{\Phi_T,\Phi_R}\Bigg[\mathbb E_{\Phi_E}\Bigg[\nonumber\\
&&\prod_{z\in \Phi_E}\mathbb E_{h_{y_0,z},\alpha,\mathsf{G}_{J,R}^{z,y_0}}\left[e^{-s\mathbf{1}_J^z\frac{P_J\mathsf{G}_{J,R}^{z,y_0} h_{y_0,z}}{d_{y_0,z}^{\alpha}}}\right]\Bigg]\Bigg]\nonumber\\
&=&\mathbb E_{\Phi_T,\Phi_R}\Bigg[\exp\Bigg(-2\pi \lambda_E\int_{0}^{\infty}\Bigg(\nonumber\\
&&\quad 1-\mathbb E_{\mathsf{G}_{J,R}^{z,y_0},h_{y_0,z},\alpha}\left[e^{-\frac{s\mathbf{1}_J^zP_J \mathsf{G}_{J,R}^{z,y_0} h_{y_0,z}}{v^{\alpha}}}\right]\Bigg)v\mathrm d v\Bigg)\Bigg]\nonumber.
\end{IEEEeqnarray}
Applying the Jensen's inequality yields
\begin{IEEEeqnarray}{rCl}
\mathcal {L}_{y_0}^J(s)
\!&\geq\!&\exp\Bigg(-2\pi \lambda_E\int_{0}^{\infty}\Bigg(1-\mathbb E_{\Phi_T,\Phi_R}\Bigg[\nonumber\\
&&\quad \mathbb E_{\mathsf{G}_{J,R}^{z,y_0},h_{y_0,z},\alpha}\left[e^{-\frac{s\mathbf{1}_J^zP_J \mathsf{G}_{J,R}^{z,y_0} h_{y_0,z}}{v^{\alpha}}}\right]\Bigg]\Bigg)v\mathrm d v\Bigg)\nonumber\\
\!&=&\!\exp\Bigg(-2\pi \lambda_E\int_{0}^{\infty}\Bigg(1-\mathbb P(\mathbf{1}_J^z=0)-\mathbb P(\mathbf{1}_J^z=1)\nonumber\\
&&\quad \mathbb E_{\mathsf{G}_{J,R}^{z,y_0},h_{y_0,z},\alpha}\left[e^{-\frac{sP_J\mathsf{G}_{J,R}^{z,y_0} h_{y_0,z}}{v^{\alpha}}}\right]\Bigg)v\mathrm d v\Bigg)\nonumber\\
\!&=&\!\exp\Bigg(-2\pi \lambda_E\int_{0}^{\infty}\mathbb P(\mathbf{1}_J^z=1)\Bigg(1-\nonumber\\
&&\quad \mathbb E_{\mathsf{G}_{J,R}^{z,y_0},h_{y_0,z},\alpha}\left[e^{-\frac{sP_J\mathsf{G}_{J,R}^{z,y_0} h_{y_0,z}}{v^{\alpha}}}\right]\Bigg)v\mathrm d v\Bigg)\nonumber\\
&=&\exp\Bigg(-2\pi \lambda_E\int_{0}^{\infty}\Omega_{J,R}(s,v)\zeta(v)v\mathrm d v\Bigg).
\end{IEEEeqnarray}
\end{IEEEproof}

\subsection{Derivation of $\mathcal{L}_{y_0}^{J}(s)$ under LOWJ Attack} \label{sec:ly0j-sal}
Similar to the analysis in Section \ref{sec:ly0j-sad}, we first establish the following lemma. 
\begin{lemma}\label{lemma:zeta-sal}
The probability that an eavesdropper $z$ with distance $v$ and link status $\tau\in\{L,N\}$ to $y_0$ is a jammer under the LOWJ attack is
\begin{IEEEeqnarray}{rCl}\label{eqn:zeta-sal}
\zeta_{\tau}(v)=\int_{\lvert v-r_0\rvert}^{v+r_0}\sum_{\kappa\in\{L,N\}}\zeta_{\kappa}^{\tau}(u,v) p_\kappa(u) f_{d_{x_0,z}}(u) \mathrm du.
\end{IEEEeqnarray}
where $\zeta_{\kappa}^{\tau}(u,v)$ is
\begin{IEEEeqnarray}{rCl}
\zeta_{\kappa}^{\tau}(u,v) = \int_{0}^{\frac{u^{\alpha_{\kappa}}}{\rho}}\!\!\!\!\!e^{-(\Lambda(\lambda, \rho  w)+\Lambda(\lambda, w))} \Lambda'(\lambda, w)\mathrm d  w
\end{IEEEeqnarray} 
for $u^{\alpha_{\kappa}}< \rho v^{\alpha_{\tau}}$ and is
\begin{IEEEeqnarray}{rCl}
\zeta_{\kappa}^{\tau}(u,v) =1-\rho \int_0^{v^{\alpha_{\tau}}}\!\!\!\!\!\! e^{-(\Lambda(\lambda, \rho  w)+\Lambda(\lambda, w))} \Lambda'(\rho w )\mathrm d  w
\end{IEEEeqnarray}
for $u^{\alpha_{\kappa}}\ge \rho v^{\alpha_{\tau}}$, where
\begin{IEEEeqnarray}{rCl}
\Lambda(\lambda, w)&=&\lambda \pi w^{\frac{2}{\alpha_N}}-\frac{2\pi\lambda}{\beta^2}(1+\beta w^{\frac{1}{\alpha_L}})e^{-\beta w^{\frac{1}{\alpha_L}}}\nonumber\\
&&+\frac{2\pi\lambda}{\beta^2}(1+\beta w^{\frac{1}{\alpha_N}})e^{-\beta w^{\frac{1}{\alpha_N}}},
\end{IEEEeqnarray}
and $\Lambda'(\lambda,  w)$ is the derivative of $\Lambda(\lambda, w)$.
\end{lemma}
\begin{IEEEproof}
See Appendix \ref{app:proof-lemma-zeta-sal}.
\end{IEEEproof}

With the help of Lemma \ref{lemma:zeta-sal}, we derive the Laplace transform of $I_{y_0}^J$ under the LOWJ attack in the following lemma.
\begin{lemma}\label{lemma:ly0j-sal}
The Laplace transform of $I_{y_0}^J$ under the LOWJ attack can be lower bounded by 
\begin{IEEEeqnarray}{rCl}\label{eqn:ly0j-sal}
\mathcal {L}_{y_0}^J(s)\!&\ge&\!\exp\!\left(\!-2\pi\lambda_E\!\int_{0}^{\infty}\!\!\!\!\!\!\sum_{\tau\in\{L,N\}}\!\!\!\!\!\Omega_{J,R}^{\tau}(s,v)\zeta_{\tau}(v)v\mathrm d v\!\right),
\end{IEEEeqnarray}
where $\Omega_{J,R}^{\tau}(s,v)$ can be obtained from \eqref{eqn:omegat1t2k}.
\end{lemma}
\begin{IEEEproof}
We divide $\Phi_E$ into independent sub-PPPs $\Phi_E^{\tau}$ of eavesdroppers with link status $\tau\in\{L,N\}$ to $y_0$, i.e., $\Phi_E=\cup_{\tau}\Phi_E^{\tau}$.
Formally, $\Phi_E^{\tau}$ is given by $\Phi_E^{\tau}=\{z\in \Phi_E: \mathcal S_{z,y_0}=\tau\}$.
Hence, we have $I_{y_0}^J=\sum_{\tau}I_{y_0}^{J,\tau}$, where $I_{y_0}^{J,\tau}=\sum_{z\in \Phi_E^{\tau}} \mathbf{1}_J^z P_J \mathsf{G}_{J,R}^{z,y_0}h_{z,y_0}^{\tau}d_{z,y_0}^{-\alpha_{\tau}}$, and thus 
\begin{IEEEeqnarray}{rCl}\label{eqn:ly0j-sal-2}
\mathcal {L}_{y_0}^J(s)&=&\prod_{\tau}\mathcal {L}_{y_0}^{J,\tau}(s),
\end{IEEEeqnarray}
where $\mathcal {L}_{y_0}^{J,\tau}(s)$ is the Laplace transform of $I_{y_0}^{J,\tau}$.
It follows from Lemma \ref{lemma:ly0j-sad} that
\begin{IEEEeqnarray}{rCl}\label{eqn:ly0jtau}
\mathcal {L}_{y_0}^{J,\tau}(s)&\ge&\exp\left(-2\pi\lambda_E\int_{0}^{\infty}\Omega_{J,R}^{\tau}(s,v)\zeta_{\tau}(v)v\mathrm d v\right).
\end{IEEEeqnarray}
Substituting \eqref{eqn:ly0jtau} into \eqref{eqn:ly0j-sal-2} completes the proof.
\end{IEEEproof}

\subsection{Derivation of $\mathcal{L}_{y_0}^{J}(s)$ under POWJ Attack} \label{sec:ly0j-sap}
The probability of being a jammer also depends on the effective antenna gain to the receiver $y_0$ under the POWJ attack and is given by the following lemma.
\begin{lemma}\label{lemma:zeta-sap}
The probability that an eavesdropper $z$ with distance $v$, link status $\tau\in\{L,N\}$ and effective antenna gain $G_n^JG_o^R$ ($n,o\in\{M,S\}$) to $y_0$ is a jammer under the POWJ attack can be approximated by 
\begin{IEEEeqnarray}{rCl}\label{eqn:zeta-sap}
\zeta_{\tau,n,o}(v)\!\!&\approx&\!\!\int_{\lvert v\!-\!r_0\rvert}^{v\!+\!r_0}\! \sum_{\kappa,l,m}\!\!p_{lm}^{TW}\!\!\zeta_{\kappa,l,m}^{\tau,n,o}(u,\!v)p_{\kappa}(u)f_{d_{x_0,z}}\!(u)\mathrm d u
\end{IEEEeqnarray}
where $\kappa \in\{L,N\}$, $l\in\{M,S\}$, $m\in\{M,S\}$, $\zeta_{\kappa,l,m}^{\tau,n,o}(u,v)$ is
\begin{IEEEeqnarray}{rCl}\label{eqn:zetakappalmtauno1}
\int_{0}^{\frac{\eta_1}{\rho}}e^{-(\hat{\Lambda}_1(\lambda, \rho w)+\hat{\Lambda}_2(\lambda, w))} \hat{\Lambda}'_2(\lambda, w)\mathrm dw
\end{IEEEeqnarray}
for $\eta_1<\rho\eta_2$ with $\eta_1=\frac{u^{\alpha_\kappa}}{P_T G_l^TG_m^W}$, $\eta_2=\frac{v^{\alpha_\tau}}{P_J G_n^JG_o^R}$, and is
\begin{IEEEeqnarray}{rCl}\label{eqn:zetakappalmtauno2}
1-\rho \int_{0}^{\eta_2}\!\!e^{-(\hat{\Lambda}_1(\lambda, \rho w)+\hat{\Lambda}_2(\lambda, w))}\hat{\Lambda}'_1(\lambda, \rho w)\mathrm dw
\end{IEEEeqnarray}
for $\eta_1\ge \rho\eta_2$, where
\begin{IEEEeqnarray}{rCl}
\hat{\Lambda}_1(\lambda, w)\!=\!\sum_{i\in\{M,S\}}\sum_{j\in\{M,S\}}\!\!\!\!\!\Lambda(p_{ij}^{TW}\lambda, P_TG_i^T G_j^W w),
\end{IEEEeqnarray}
\begin{IEEEeqnarray}{rCl}
\hat{\Lambda}_2(\lambda, w)\!=\!\sum_{i\in\{M,S\}}\sum_{j\in\{M,S\}}\!\!\!\!\!\Lambda(p_{ij}^{JR}\lambda, P_J G_i^JG_j^R w),
\end{IEEEeqnarray}
$\hat{\Lambda}'_1(\lambda, w)$ and $\hat{\Lambda}'_2(\lambda, w)$ are the derivatives of $\hat{\Lambda}_1(\lambda, w)$ and $\hat{\Lambda}_2(\lambda, w)$, respectively.
\end{lemma}
\begin{IEEEproof}
See Appendix \ref{app:proof-lemma-zeta-sap}.
\end{IEEEproof}
Note that the approximation in \eqref{eqn:zeta-sap} is due to the fact that we neglect the dependence between $\mathsf{G}_{T,W}^{x_0,z}$ and $\mathsf{G}_{J,R}^{z,y_0}$.
Given the probability $\zeta_{\tau,n,o}(v)$, we derive the Laplace transform of $I_{y_0}^J$ under the POWJ attack as follows.
\begin{lemma}\label{lemma:ly0j-sap}
The Laplace transform of $I_{y_0}^J$ under the POWJ attack can be approximated by 
\begin{IEEEeqnarray}{rCl}\label{eqn:ly0j-sap-2}
\mathcal {L}_{y_0}^J(s)\!&\approx&\!\exp\!\left(\!\!-2\pi\lambda_E\!\!\int_{0}^{\infty}\!\!\!\!\sum_{\tau,n,o}\!\!\Omega_{J,R}^{\tau,n,o}(s,v)\zeta_{\tau,n,o}(v)v\mathrm d v\!\!\right),
\end{IEEEeqnarray}
where $\Omega_{J,R}^{\tau,n,o}(s,v)$ can be obtained from \eqref{eqn:omegat1t2kij}.
\end{lemma}
\begin{IEEEproof}
Similar to the proof of Lemma \ref{lemma:ly0j-sal}, $\mathcal {L}_{y_0}^J(s)$ can be rewritten as
\begin{IEEEeqnarray}{rCl}\label{eqn:ly0j-sap-2}
\mathcal {L}_{y_0}^J(s)&=&\prod_{\tau,n,o}\mathcal {L}_{y_0}^{J,\tau,n,o}(s),
\end{IEEEeqnarray}
where $\mathcal {L}_{y_0}^{J,\tau,n,o}(s)$ is the  the Laplace transform of the interference caused by the eavesdroppers with link status $\tau$ and effective antenna gain $G_n^JG_o^R$ to $y_0$.
Based on  Lemma \ref{lemma:ly0j-sad}, $\mathcal {L}_{y_0}^{J,\tau,n,o}(s)$ can be given by
\begin{IEEEeqnarray}{rCl}\label{eqn:ly0jtauno}
\mathcal {L}_{y_0}^{J,\tau,n,o}(s)\!&\ge&\!\exp\!\left(\!\!-2\pi\lambda_E\int_{0}^{\infty}\!\!\Omega_{J,R}^{\tau,n,o}(s,v)\zeta_{\tau,n,o}(v)v\mathrm d v\!\!\right).
\end{IEEEeqnarray}
Substituting \eqref{eqn:ly0jtauno} into \eqref{eqn:ly0j-sap-2} completes the proof.
\end{IEEEproof}

\section{Secrecy Probability Analysis}\label{sec:sp-analysis}
In this section, we focus on the typical link $x_0\rightarrow y_0$ again and derive a unified expression of the secrecy probability. We then analyze the key term involved in the unified expression, i.e., the Laplace transforms of the interference from the concurrent transmitters to any eavesdropper under the DOWJ, LOWJ and POWJ attacks in Sections \ref{sec:lzt-sad}, \ref{sec:lzt-sal} and \ref{sec:lzt-sap}, respectively.

\subsection{Secrecy Probability}\label{sec:ps-general}
The secrecy probability is formulated as 
\begin{IEEEeqnarray}{rCl}
p_s=\mathbb P\left(\bigcap_{z\in\Phi_W}\{\log(1+\mathrm{SINR}_{x_0,z})\leq R_e\}\right),
\end{IEEEeqnarray}
where $\mathrm{SINR}_{x_0,z}$ denotes the SINR of a wiretapper $z\in\Phi_W$.
We consider an equivalent formulation of $p_s$ by assuming that all eavesdroppers wiretap on the typical link.
Since jammers actually do not wiretap, we remove their impact by setting their interferences to infinity.
We divide the PPP $\Phi_E$ into sub-PPPs $\Phi_E^{\kappa, l,m}$ of eavesdroppers with link status $\kappa\in \{L,N\}$ and antenna gain $G_l^TG_m^W$ ($l, m\in\{M,S\}$) to $x_0$.
Hence, $p_s$ can be rewritten as
\begin{IEEEeqnarray}{rCl}\label{eqn:ps-formulation}
p_s&=&\prod_{\kappa,l,m}p_s^{\kappa,l,m}\\
&=&\prod_{\kappa,l,m}\mathbb P\left(\bigcap_{z\in\Phi_E^{\kappa,l,m}}\{\log(1+\mathrm{SINR}_{x_0,z}^{\kappa,l,m})\leq R_e\}\right).\nonumber
\end{IEEEeqnarray}

We assume that eavesdroppers can eliminate the interference from the jammers. 
Thus, for any eavesdropper $z\in \Phi_E^{\kappa,l,m}$, we have
\begin{IEEEeqnarray}{rCl}
\mathrm{SINR}_{x_0,z}^{\kappa,l,m}=\frac{P_T \mathsf{G}_{T,W}^{x_0,z} h_{x_0,z}d_{x_0,z}^{-\alpha}}{\hat{I}_z^T+\sigma^2},
\end{IEEEeqnarray}
where
\begin{align} 
\hat{I}_z^T =
  \begin{cases}
   I_z^T, &\mathbf{1}_J^z=0 \\
   \infty,  &\mathbf{1}_J^z=1\\
  \end{cases}
\end{align}
denotes the interference from concurrent transmitters. Here, $I_z^T=\sum_{x\in\Phi_T\backslash\{x_0\}} P_T\mathsf{G}_{T,W}^{x,z} h_{x,z}d_{x,z}^{-\alpha}$.
We now derive the secrecy probability based on the above formulations.
\begin{theorem}\label{theorem:ps}
The secrecy probability of the typical transmission pair $x_0\rightarrow y_0$ under the DOWJ, LOWJ and POWJ attacks can be approximated by 
\begin{IEEEeqnarray}{rCl}\label{eqn:ps}
p_s&\approx&\exp\Bigg(-2\pi \lambda_E\sum_{\kappa,l,m} p_{lm}^{TW}\sum_{t=1}^{N_{\kappa}}\binom{N_{\kappa}}{t}(-1)^{t+1}\\
&&\int_0^\infty p_{\kappa}(u) e^{-t\nu_{\kappa,l,m} u^{\alpha_{\kappa}}\sigma^2}\mathcal{L}_{z,\kappa,l,m}^{T}(t\nu_{\kappa,l,m} u^{\alpha_{\kappa}}, u)u\mathrm du\Bigg),\nonumber
\end{IEEEeqnarray}
where $\nu_{\kappa,l,m}= \frac{N_{\kappa}(N_{\kappa}!)^{-1/N_{\kappa}}(2^{R_e}-1)}{P_TG_l^TG_m^W}$ and $\mathcal L_{z,
\kappa,l,m}^{T}(\cdot,u)$ denotes the Laplace transform of $\hat{I}_z^T$ for any eavesdropper $z\in \Phi_E^{\kappa,l,m}$ with $d_{x_0,z}=u$. 
\end{theorem}
\begin{IEEEproof}
According to \eqref{eqn:ps-formulation}, we first derive $p_s^{\kappa,l,m}$, which is given by
\begin{IEEEeqnarray}{rCl}\label{eqn:ps-proof}
p_s^{\kappa,l,m}&=&\mathbb P\left(\bigcap_{z\in\Phi_E^{\kappa,l,m}}\{\log(1+\mathrm{SINR}_{x_0,z}^{\kappa,l,m})\leq R_e\}\right)\\
&\approx&\exp\Bigg(-2\pi p_{lm}^{TW}\lambda_E\sum_{t=1}^{N_{\kappa}}\binom{N_{\kappa}}{t}(-1)^{t+1}\nonumber\\
&&\int_0^\infty p_{\kappa}(u) e^{-t\nu_{\kappa,l,m} u^{\alpha_{\kappa}}\sigma^2}\mathcal{L}_{z}^{T}(t\nu_{\kappa,l,m} u^{\alpha_{\kappa}}, u)u\mathrm du\Bigg)\nonumber,
\end{IEEEeqnarray}
following from Theorem $2$ in \cite{YZhang2022TIFS}.
Substituting \eqref{eqn:ps-proof} into \eqref{eqn:ps-formulation} completes the proof.
\end{IEEEproof}

From \eqref{eqn:ps}, we know that the key term involved in $p_s$ is the Laplace transform $\mathcal{L}_{z,\kappa,l,m}^{T}(s, u)$.
In what follows, we will derive the expressions of $\mathcal{L}_{z,\kappa,l,m}^{T}(s, u)$ under the DOWJ, LOWJ and POWJ attacks, respectively. 
Prior to the derivation, we first give the following functions for any $k\in\{L,N\}$, $i\in\{M,S\}$, $j\in\{M,S\}$, $t_1\in \{T,J\}$ and  $t_2\in\{R,W\}$.
\begin{IEEEeqnarray}{rCl}\label{eqn:xit1t2k,i,j}
\Xi_{t_1,t_2}^{k,i,j}(s,\lambda,c)&=&\exp\left(-2\pi \lambda \int_{c}^{\infty}\Omega_{t_1,t_2}^{k,i,j}(s,r)r \mathrm dr \right),
\end{IEEEeqnarray}
\begin{IEEEeqnarray}{rCl}\label{eqn:xit1t2k}
\Xi_{t_1,t_2}^{k}(s,\lambda,c)&=&\prod_{i,j}\Xi_{t_1,t_2}^{k,i,j}(s,\lambda,c)\nonumber\\
&=&\exp\left(-2\pi \lambda \int_{c}^{\infty}\Omega_{t_1,t_2}^{k}(s,r)r \mathrm dr \right),
\end{IEEEeqnarray}
\begin{IEEEeqnarray}{rCl}\label{eqn:xit1t2}
\Xi_{t_1,t_2}(s,\lambda,c)&=&\prod_{k}\Xi_{t_1,t_2}^{k}(s,\lambda,c)\nonumber\\
&=&\exp\left(-2\pi \lambda \int_{c}^{\infty}\Omega_{t_1,t_2}(s,r)r \mathrm dr \right),
\end{IEEEeqnarray}
where $\Omega_{t_1,t_2}^{k,i,j}(s,r)$, $\Omega_{t_1,t_2}^{k}(s,r)$ and $\Omega_{t_1,t_2}(s,r)$ are given in \eqref{eqn:omegat1t2kij}, \eqref{eqn:omegat1t2k} and \eqref{eqn:omegat1t2}, respectively.


\subsection{Derivation of $\mathcal{L}_{z,\kappa,l,m}^{T}(s, u)$ under DOWJ attack}\label{sec:lzt-sad}
We can see from Section \ref{sec:sa-strategy} that the DOWJ attack is dependent solely on the distances from $z$ to the transmitters and receivers, which means that the Laplace transform $\mathcal{L}_{z,\kappa,l,m}^{T}(s, u)$ varies with only $d_{x_0,z}=u$.
Thus, we rewrite $\mathcal{L}_{z,\kappa,l,m}^{T}(s, u)$ as $\mathcal{L}_{z}^{T}(s, u)$ under the DOWJ attack, whose expression can be given in the following lemma.
\begin{lemma}\label{lemma:ltzt-sad}
The Laplace transform of the interference caused by the concurrent transmitters at any eavesdropper $z \in \Phi_E^{\kappa,l,m}$ with distance $u$ to the typical transmitter $x_0$ under the DOWJ attack is given by 
\begin{IEEEeqnarray}{rCl}
\mathcal L_z^T(s,u)&=&\Xi_{T,W}(s,\lambda,0)-\bar{F}_{d_{y_0,z}}(c_u)\\
&&\times\int_{0}^{u/\rho}\Xi_{T,W}(s,\lambda,\rho  w )2\pi\lambda  w e^{-(\rho^2+1)\lambda \pi w ^2} \mathrm d   w \nonumber\\
&&-\int_{\lvert u-r_0 \rvert}^{c_u}\Xi_{T,W}(s,\lambda,\rho v)e^{-(\rho^2+1)\lambda\pi v^2}f_{d_{y_0,z}}(v)\mathrm d v\nonumber\\
&&-\int_{\lvert u-r_0 \rvert}^{c_u}\int_{0}^{v}\Xi_{T,W}(s,\lambda,\rho  w )\nonumber\\
&&\quad \quad \quad \quad \quad 2\pi\lambda  w e^{-(\rho^2+1)\lambda \pi w ^2}f_{d_{y_0,z}}(v)\mathrm d   w \mathrm d v\nonumber,
\end{IEEEeqnarray}
where $c_u=\min\{\max\{\frac{u}{\rho}, \lvert u-r_0\rvert\},u+r_0\}$,
\begin{IEEEeqnarray}{rCl}\label{eqn:ccdf_dy0z}
\bar{F}_{d_{y_0,z}}(v)&=&1-\frac{1}{\pi}\arccos\left(\frac{r_0^2+u^2-v^2}{2r_0u}\right)
\end{IEEEeqnarray}
is the CCDF of $d_{y_0,z}$ and
\begin{IEEEeqnarray}{rCl}\label{eqn:pdf_dy0z}
f_{d_{y_0,z}}(v)&=&\frac{2v}{\pi\sqrt{4r_0^2u^2-(r_0^2+u^2-v^2)^2}}
\end{IEEEeqnarray}
is the corresponding PDF.
\end{lemma}
\begin{IEEEproof}
See Appendix \ref{app:lemma-ltzt-sad}.
\end{IEEEproof}

\begin{figure*}[!t]
  \normalsize
\begin{IEEEeqnarray}{rCl}\label{eqn:ltztkappatau}
\mathcal L_{z,\kappa}^{T,\tau}(s,u,v)=
\begin{dcases}
	\Xi_{T,W}(s,\lambda, 0)-\int_{0}^{\frac{u^{\alpha_{\kappa}}}{\rho}}\!\! e^{-(\Lambda(\lambda, \rho  w)+\Lambda(\lambda, w))}\Lambda'(\lambda, w)\!\! \prod\limits_{k\in\{L,N\}}\!\!\Xi_{T,W}^{k}\left(s,\lambda, (\rho w)^{\frac{1}{\alpha_{k} }}\right)\mathrm dw, &\!\! u^{\alpha_{\kappa}}<\rho v^{\alpha_{\tau}}\\
	\\
	\Xi_{T,W}(s,\lambda, 0)-e^{-(\Lambda(\lambda, \rho v^{\alpha_{\tau}})+\Lambda(\lambda, v^{\alpha_{\tau}}))}\!\! \prod\limits_{k\in\{L,N\}}\!\!\Xi_{T,W}^{k}\left(s,\lambda,  (\rho 	v^{\alpha_{\tau}})^{\frac{1}{\alpha_{k}}}\right)\\
 	-\int_0^{v^{\alpha_{\tau}}}e^{-(\Lambda(\lambda, \rho  w)+\Lambda(\lambda, w))}\Lambda'(\lambda, w)\!\! \prod\limits_{k\in\{L,N\}}\!\!\Xi_{T,W}^{k}\left(s,\lambda, (\rho 		w)^{\frac{1}{\alpha_{k} }}\right)\mathrm d  w, &\!\! u^{\alpha_{\kappa}}\geq \rho v^{\alpha_{\tau}}
\end{dcases}
\end{IEEEeqnarray}
	\vspace*{4pt}
\end{figure*}
\begin{figure*}[!t]
  \normalsize
\begin{IEEEeqnarray}{rCl}\label{eqn:ltztkappalmtauno}
\mathcal L_{z,\kappa,l,m}^{T,\tau,n,o}(s,u,v)\approx
\begin{dcases}
	\Xi_{T,W}(s,\lambda,0)\!-\!\int_{0}^{\frac{\eta_1}{\rho}}\!\!\!\!e^{-\left(\hat{\Lambda}_1(\lambda, \rho w)+\hat{\Lambda}_2(\lambda, w)\right)}\hat{\Lambda}'_2(\lambda, w)\! \prod_{k,i,j}\!\Xi_{T,W}^{k,i,j}\left(s,\lambda,(\rho  w P_T G_i^T G_j^W)^{\frac{1}{\alpha_k}}\right)\!\mathrm dw , \!\!&\!\! \eta_1<\rho \eta_2\\
	\\
	\Xi_{T,W}(s,\lambda, 0)\!-\!e^{-\left(\hat{\Lambda}_1(\lambda, \rho\eta_2)+\hat{\Lambda}_{2}(\lambda,\eta_2)\right)}\!\prod_{k,i,j}\!\Xi_{T,W}^{k,i,j}\left(s,\lambda,  (\rho\eta_2 P_T G_i^T G_j^W)^{\frac{1}{\alpha_{k}}}\right) \\
	-\int_0^{\eta_2}\!\!\!\!e^{-(\hat{\Lambda}_1(\lambda, \rho w)+\hat{\Lambda}_2(\lambda, w))}\hat{\Lambda}'_2(\lambda, w) \!\prod_{k,i,j}\!\Xi_{T,W}^{k,i,j}\left(s,\lambda,(\rho  w P_T G_i^T G_j^W)^{\frac{1}{\alpha_k}}\right)\mathrm d  w, \!\!&\!\! \eta_1\geq \rho \eta_2
\end{dcases}
\end{IEEEeqnarray}
	\vspace*{4pt}
\end{figure*}

\subsection{Derivation of $\mathcal{L}_{z,\kappa,l,m}^{T}(s, u)$ under LOWJ attack}\label{sec:lzt-sal}
The LOWJ attack depends on both the distances and link status from $z$ to the transmitters and receivers.
Thus, the Laplace transform $\mathcal{L}_{z,\kappa,l,m}^{T}(s, u)$ under the LOWJ attack varies with both $d_{x_0,z}=u$ and $\mathcal S_{x_0,z}=\kappa$.
Rewriting $\mathcal{L}_{z,\kappa,l,m}^{T}(s, u)$ as $\mathcal{L}_{z,\kappa}^{T}(s, u)$, we give the Laplace transform $\mathcal{L}_{z,\kappa}^{T}(s, u)$ in the following lemma.
\begin{lemma}\label{lemma:ltzt-sal}
The Laplace transform of the interference caused by the concurrent transmitters at any eavesdropper $z \in \Phi_E^{\kappa,l,m}$ with distance $u$ and link status $\kappa$ to the typical transmitter $x_0$ under the LOWJ attack is
\begin{IEEEeqnarray}{rCl}
\mathcal L_{z,\kappa}^T(s,u)\!&=&\!\int_{\lvert u\!-\!r_0\rvert}^{u\!+\!r_0}\!\!\sum_{\tau}\mathcal L_{z,\kappa}^{T,\tau}(s,u,v)p_\tau(v) f_{d_{y_0,z}}(v)\mathrm dv,
\end{IEEEeqnarray}
where $\mathcal L_{z,\kappa}^{T,\tau}(s,u,v)$ is given by \eqref{eqn:ltztkappatau}.
\end{lemma}
\begin{IEEEproof}
See Appendix \ref{app:lemma-ltzt-sal}.
\end{IEEEproof}

\subsection{Derivation of $\mathcal{L}_{z,\kappa,l,m}^{T}(s, u)$ under POWJ attack}\label{sec:lzt-sap}
Note that the Laplace transform $\mathcal{L}_{z,\kappa,l,m}^{T}(s, u)$ under the POWJ attack varies with  $d_{x_0,z}=u$, $\mathcal S_{x_0,z}=\kappa$ and $\mathsf{G}_{T,W}^{x_0,z}=G_l^TG_m^W$.
The following lemma summarizes the Laplace transform $\mathcal{L}_{z,\kappa,l,m}^{T}(s, u)$ under the POWJ attack.
\begin{lemma}\label{lemma:ltzt-sap}
The Laplace transform of the interference caused by the concurrent transmitters at any eavesdropper $z \in \Phi_E^{\kappa,l,m}$ with distance $u$, link status $\kappa$ and antenna gain $G_l^TG_m^W$ to the typical transmitter $x_0$ under the POWJ attack can be approximated by 
\begin{IEEEeqnarray}{rCl}\label{eqn:ltzt-sap}
\mathcal L_{z,\kappa,l,m}^{T}(s,u)&\approx&\int_{\lvert u-r_0\rvert}^{u+r_0}\sum_{\tau,n,o}p_{no}^{JR}\mathcal L_{z,\kappa,l,m}^{T,\tau,n,o}(s,u,v)\nonumber\\
&&\quad \quad \quad \quad \quad p_{\tau}(v)f_{d_{y_0,z}}(v)\mathrm dv,
\end{IEEEeqnarray}
where $\mathcal L_{z,\kappa,l,m}^{T,\tau,n,o}(s,u,v)$ is given by \eqref{eqn:ltztkappalmtauno}.
\end{lemma}
\begin{IEEEproof}
See Appendix \ref{app:lemma-ltzt-sap}.
\end{IEEEproof}

\section{Numerical Results} \label{sec:num-res}  
In this section, we provide simulation results to validate the derived secrecy and connection probabilities, followed by discussions on the impacts of system parameters on the STC performance. We also compare the attack effect of the three OWJ attacks in terms of the STC performance.
\begin{table}[h]
\renewcommand{\arraystretch}{1.5}
\caption{Parameters used in simulations.}
\label{tb:pmts}
\centering
\begin{tabular}{|l|l|}
\hline
\bfseries Parameters &  \bfseries Value \\
\hline
Carrier frequency & $28$ GHz\\
Link distance $r_0$ & $50m$ \\
Channel bandwidth & $1$ GHz \\
Noise spectral density & $-174$ dBm/Hz\\
Transmit power $P_T$ & $1$ W (i.e., $30$ dBm)\\
Path loss exponent $\alpha_L$ ($\alpha_N$) & 2 (4)\\
Nakagami fading parameter $N_L$ ($N_N$) & 3 (2)\\
Block density $\beta$ & $1/141.4$\\
ML beam width $\theta_T$, $\theta_R$, $\theta_J$, $\theta_W$ &$\pi/6$\\
ML gain $G_M^T$, $G_M^R$, $G_M^E$, $G_M^J$ & $10$ \\
SL gain $G_S^T$, $G_S^R$, $G_S^E$, $G_S^J$ & $0.1$ \\
\hline
\end{tabular}
\end{table}

\subsection{Simulation and Validation}
A dedicated simulator was developed to simulate the transmission process in a Poisson mmWave bipolar network.
Using this simulator, we conducted simulations for the secrecy probability and connection probability of the network under the DOWJ, LOWJ and POWJ attacks for the settings of $\lambda=0.0001~m^{-2}$, $\lambda_E=0.0001~m^{-2}$, $P_J=10$ W, $R_t=4$ bps/Hz and $R_e=2$ bps/Hz.
The other parameters are summarized in Table \ref{tb:pmts}.

\begin{figure}[h]
 \centering
 \includegraphics[width=0.45\textwidth]{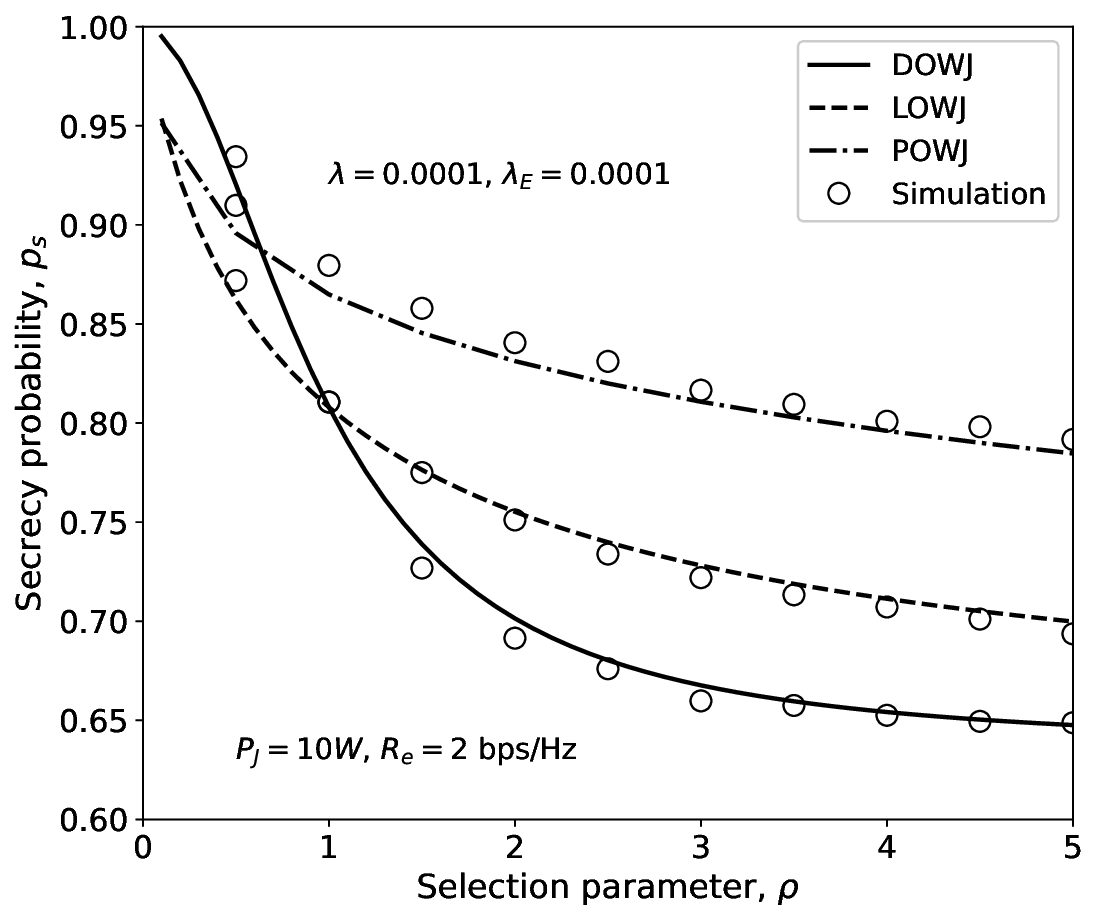}
 \caption{Secrecy probability vs. bias factor $\rho$.}
 \label{fig:psvalidation}
\end{figure}

\begin{figure}[h]
 \centering
 \includegraphics[width=0.45\textwidth]{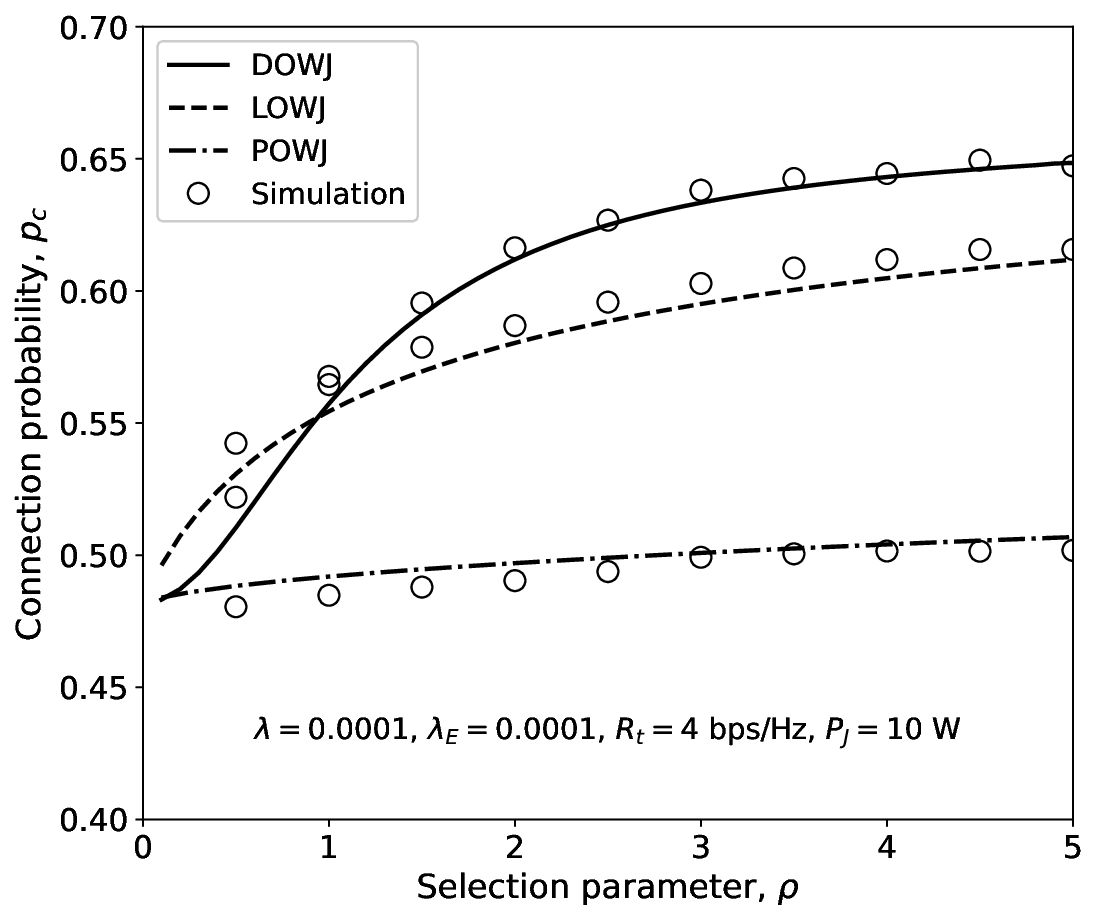}
 \caption{Connection probability vs. bias factor $\rho$.}
 \label{fig:pcvalidation}
\end{figure}

We summarize the simulation results and also the theoretical ones in Figs. \ref{fig:psvalidation} and \ref{fig:pcvalidation}.
We can see from the figures that the theoretical results provide good approximations or tight bounds for the secrecy probability and connection probability under all three OWJ attacks, implying the effectiveness of the derived analytical expressions.
We can also see from the figures that, as the bias factor $\rho$ (i.e.,  the preference for the wiretapping attack) increases, the secrecy probability decreases while the connection probability increases under all three attacks.
This is intuitive since a larger $\rho$ leads to more wiretappers and thus fewer jammers in the network.

\subsection{STC Performance Evaluation}
\subsubsection{STC vs. $\lambda_E$}
\begin{figure}[h] 
 \centering
 \includegraphics[width=0.45\textwidth]{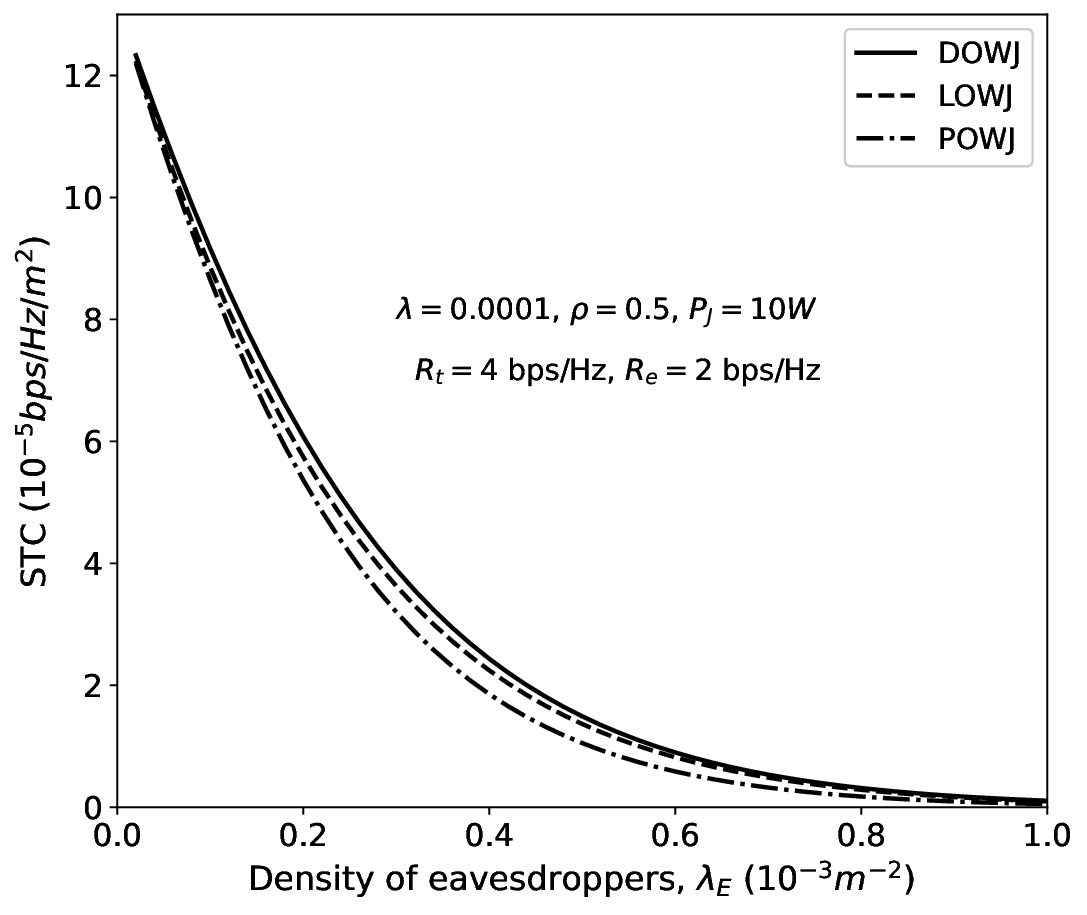}
 \caption{STC vs. eavesdropper density $\lambda_E$.}
 \label{fig:stcvsLE}
\end{figure}

We first explore the impact of the eavesdropper density $\lambda_E$ on the network STC performance, for which we show in Fig. \ref{fig:stcvsLE} STC vs. $\lambda_E$ under all the three OWJ attacks for the settings of $\lambda=0.0001$, $\rho=0.5$, $P_J=10$ W, $R_t=4$ bps/Hz and $R_e=2$ bps/Hz.
The results show that the STC decreases as $\lambda_E$ increases for a given $\rho$ under all the three OWJ attacks, which is due to the more wiretappers and jammers resulting from the increased $\lambda_E$.

\begin{figure}[h]
 \centering
 \includegraphics[width=0.45\textwidth]{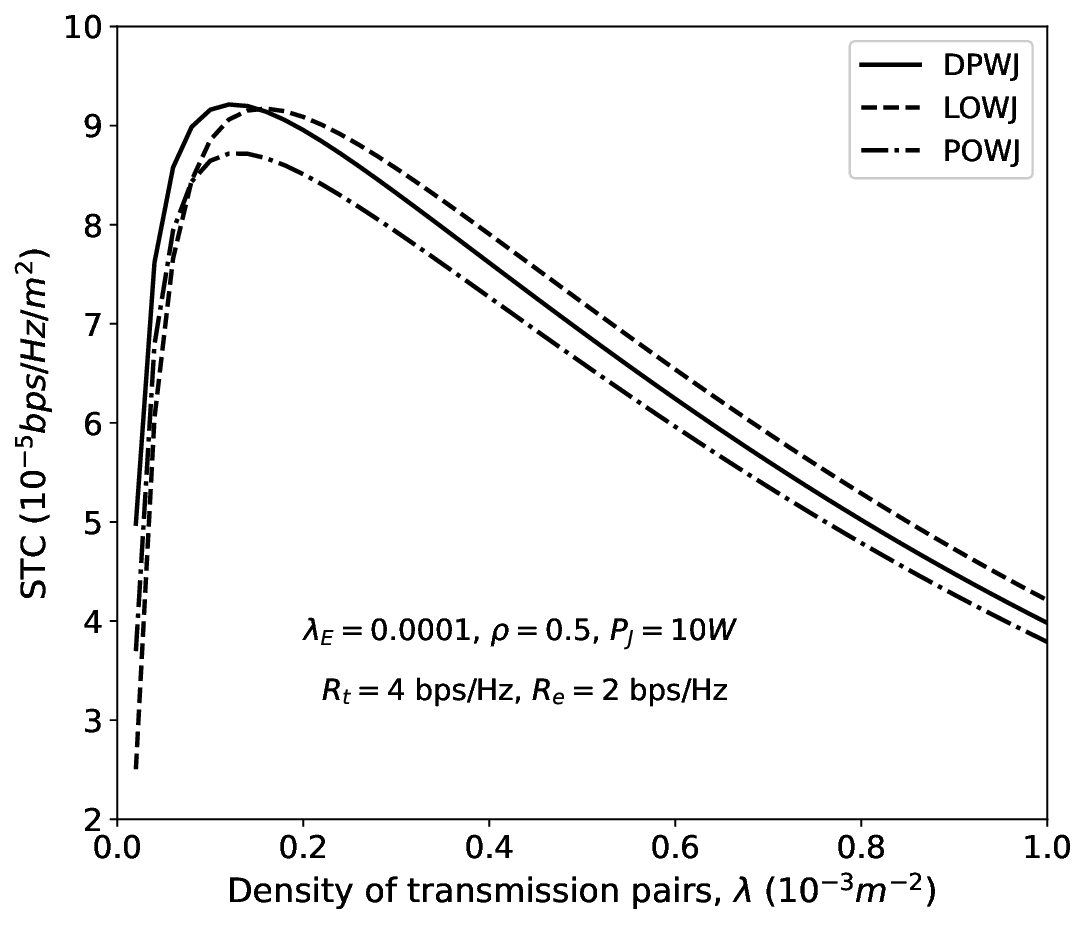}
 \caption{STC vs. transmission density $\lambda$.}
 \label{fig:stcvslt}
\end{figure}

\subsubsection{STC vs. $\lambda$}
Next, we investigate how the density of transmission pairs (i.e., $\lambda$) affects the network STC performance. Fig.~\ref{fig:stcvslt} plots the STC vs. $\lambda$ under all the three OWJ attacks for the settings of $\lambda_E=0.0001$, $\rho=0.5$, $P_J=10$ W, $R_t=4$ bps/Hz and $R_e=2$ bps/Hz.
We can see from Fig. \ref{fig:stcvslt} that, as $\lambda$ increases, the STC first increases and then decreases under all three OWJ attacks.
The reason is that the increase of $\lambda$ dominates the trend of the STC for small $\lambda$'s, while as $\lambda$ continues to increase, the secrecy probability remains almost unchanged and the decrease of the connection probability becomes the dominant factor, leading to the decrease of the STC.
The results in Fig. \ref{fig:stcvslt} reveal the existence of the optimal density of transmission pairs given a network and its key parameters, which should be taken into consideration during the network design.

\begin{figure}[h]
 \centering
 \includegraphics[width=0.45\textwidth]{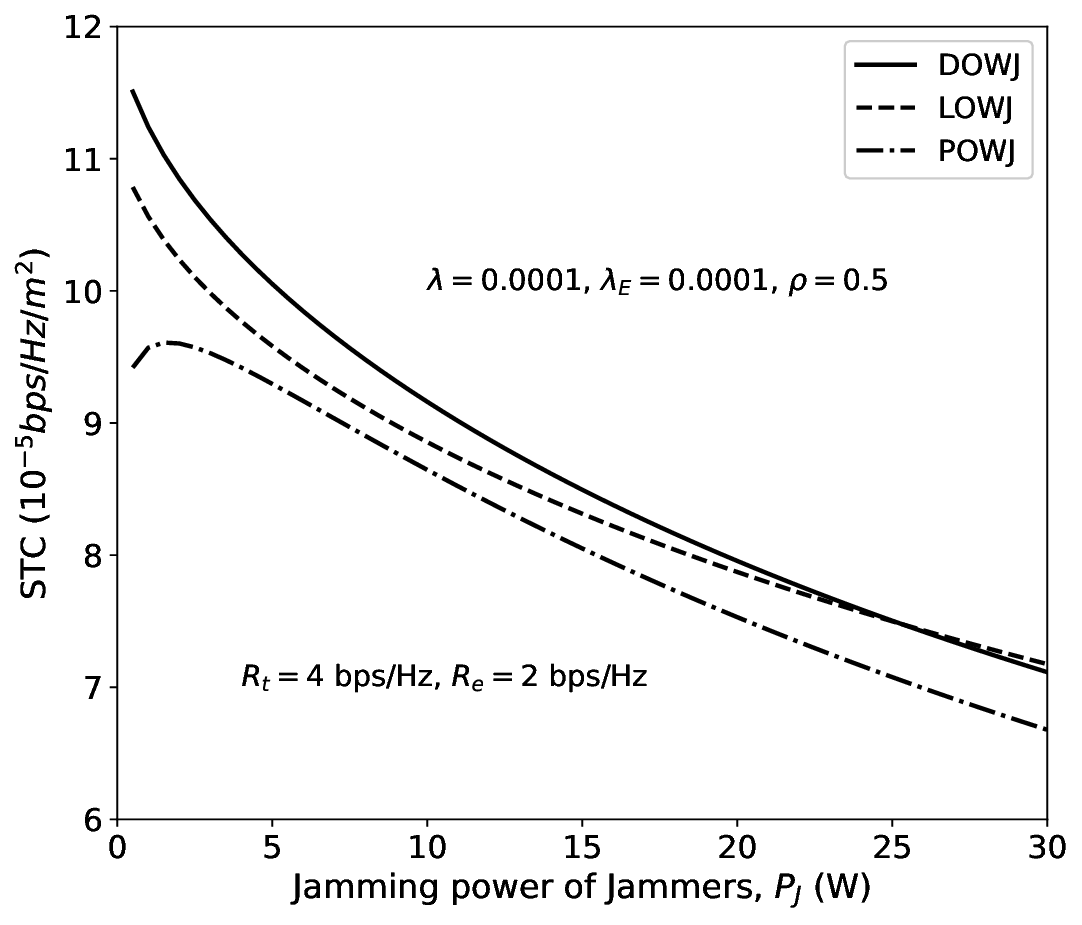}
 \caption{STC vs. jamming power $P_J$.}
 \label{fig:stcvspj}
\end{figure}
\begin{figure}[h]
 \centering
 \includegraphics[width=0.45\textwidth]{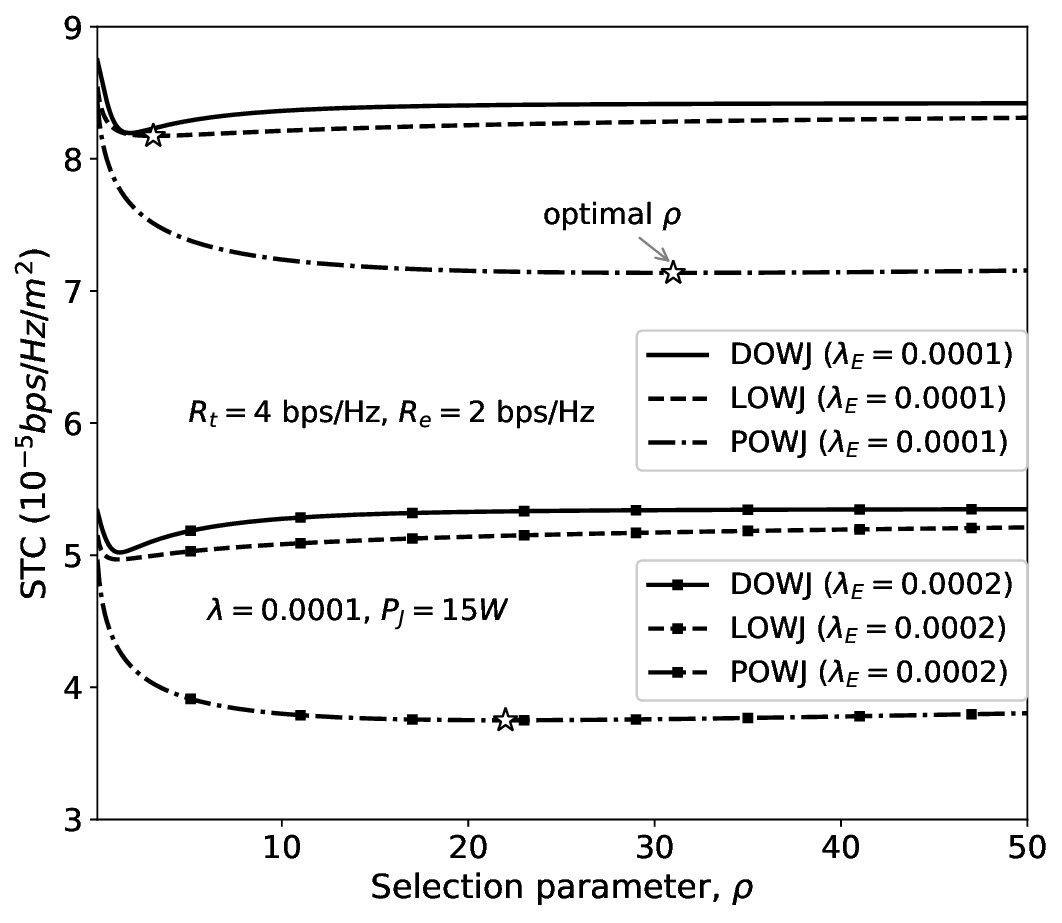}
 \caption{STC vs. bias factor $\rho$ under different settings of $\lambda_E$.}
 \label{fig:stcvsrho-le}
\end{figure}

\subsubsection{STC vs. $P_J$} We then investigate the impact of the jamming power $P_J$ on the network STC performance in Fig.~\ref{fig:stcvspj}, which plots STC vs. $P_J$ under all the three OWJ attacks for the settings of $\lambda=0.0001$, $\lambda_E=0.0001$, $\rho=0.5$, $R_t=4$ bps/Hz and $R_e=2$ bps/Hz.
We can observe from Fig.~\ref{fig:stcvspj} that, as $P_J$ increases, the STC under the DOWJ and LOWJ attacks decreases while that under the POWJ attack first increases and then decreases.
This is because the selections of attack patterns in the DOWJ and LOWJ attacks are independent of $P_J$ and thus the increase of $P_J$ leads to only the increase of the interference level to the receivers, decreasing the connection probabilities.
For the POWJ attack, as $P_J$ increases, the probability of wiretapping decreases while that of jamming increases, leading to increased secrecy probability and decreased connection probability.
The STC is dominated by the secrecy probability for small $\rho$'s and dominated by the connection probability for large $\rho$'s.

\begin{figure}[h]
 \centering
 \includegraphics[width=0.45\textwidth]{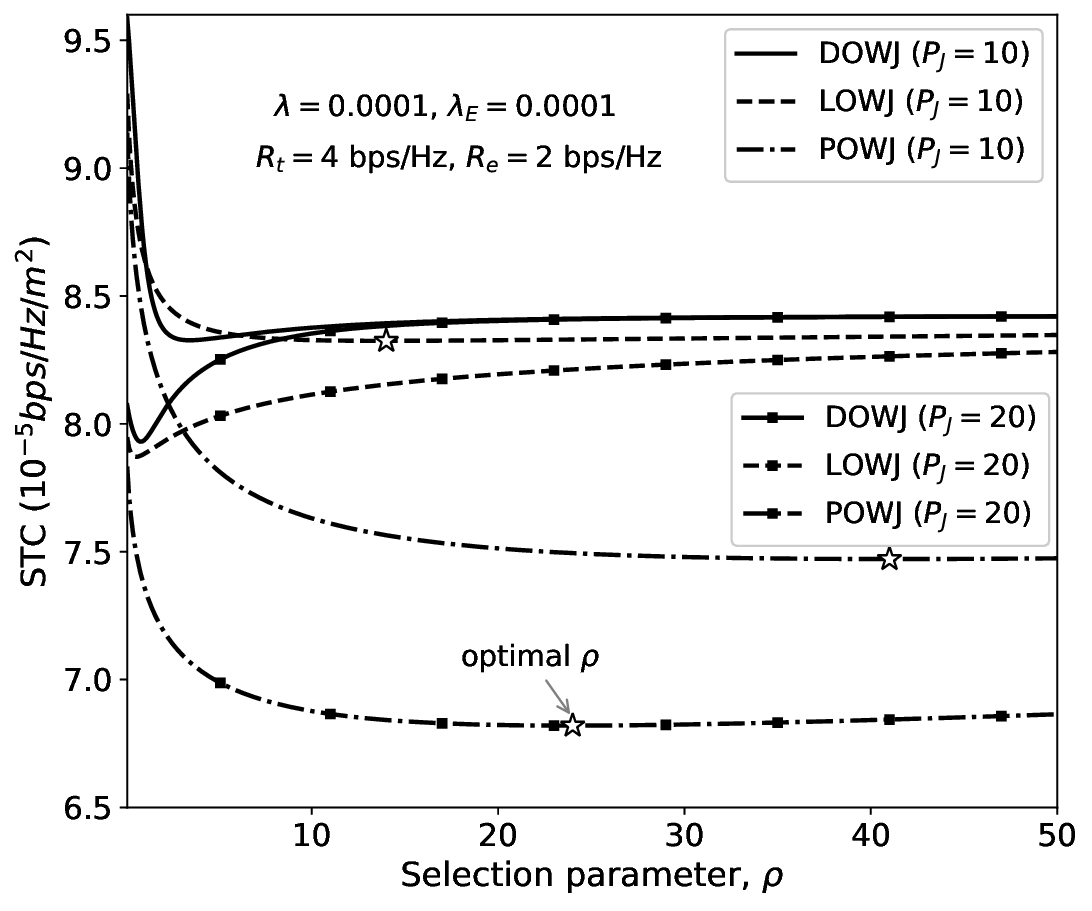}
 \caption{STC vs. bias factor $\rho$ under different settings of $P_J$.}
 \label{fig:stcvsrho-pj}
\end{figure}
\begin{figure}[h]
 \centering
 \includegraphics[width=0.45\textwidth]{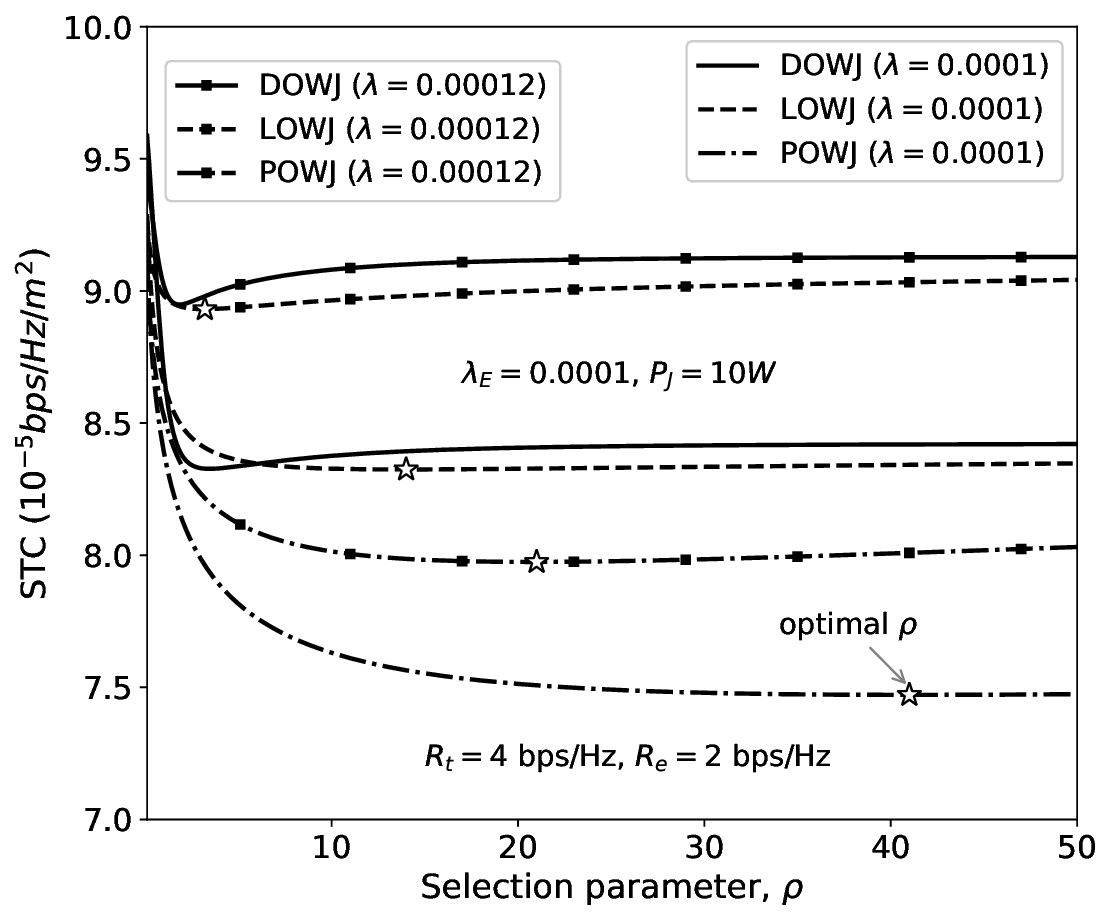}
 \caption{STC vs. bias factor $\rho$ under different settings of $\lambda$.}
 \label{fig:stcvsrho-lt}
\end{figure}

\subsubsection{STC vs. $\rho$}
We finally explore the impact of the bias factor $\rho$ on the network STC performance under the three OWJ attacks.
Fig. \ref{fig:stcvsrho-le} shows the results of STC vs. $\rho$ under the settings of $R_t=4$ bps/Hz, $R_e=2$ bps/Hz, $\lambda=0.0001$ and $P_J=15$ W.
In addition, we consider two different settings of $\lambda_E$ (i.e., $\lambda_E=0.0001$ and $\lambda_E=0.0002$).
We can see from Fig. \ref{fig:stcvsrho-le} that, as $\rho$ increases, the STC first decreases and then increases under all the three OWJ attacks, implying the existence of the optimal $\rho$ for eavesdroppers to minimize the network STC performance, i.e., maximizing the attack effect.
This shows that neither pure jamming (i.e., $\rho \rightarrow 0$) nor pure wiretapping (i.e., $\rho \rightarrow \infty$) is the optimal strategy and the OWJ attacks are more favorable for eavesdroppers.

Careful observation shows that the optimal $\rho$ decreases as $\lambda_E$ increases, indicating that eavesdroppers prefer the jamming attack more as their density increases.
To compare the attack effect of the three OWJ attacks, we focus on the worst STC performance they can achieve.
Fig. \ref{fig:stcvsrho-le} shows that the worst STC achieved by the POWJ attack is the smallest among the three OWJ attacks, which implies that the POWJ attack represents the most hazardous attack and the information of effective antenna gain plays a significant role in improving the attack effect.
We can also see that the worst STC achieved by the LOWJ attack is almost the same as that achieved by the DOWJ attack, which indicates that the information of link status has little impact on improving the attack effect.

To show the generality of our findings, we also plot STC vs. $\rho$ in Fig. \ref{fig:stcvsrho-pj} under the settings of  $\lambda=0.0001$,  $\lambda_E=0.0001$, $R_t=4$ bps/Hz, $R_e=2$ bps/Hz and two different jamming powers, i.e., $P_J=10$ W and $P_J=20$ W, and in Fig. \ref{fig:stcvsrho-lt} under the settings of $\lambda_E=0.0001$, $R_t=4$ bps/Hz, $R_e=2$ bps/Hz, $P_J=10$ W and two different densities of transmission pairs, i.e., $\lambda=0.0001$ and $\lambda=0.00012$.
Findings similar to those in Fig. \ref{fig:stcvsrho-le} can be observed from both figures.
We can also see from Fig. \ref{fig:stcvsrho-pj} and Fig. \ref{fig:stcvsrho-lt} that the optimal $\rho$ decreases as the jamming power $P_J$ and the transmission density $\lambda$ increase, respectively,  suggesting that eavesdroppers prefer the jamming attack more if they can choose a larger jamming power, or when more transmissions exist in the network.

\section{Conclusions} \label{sec:con} 
This paper proposed a new opportunistic wiretapping and jamming (OWJ) attack model for millimeter-wave (mmWave) wireless networks and provided three realizations, namely DOWJ, LOWJ and POWJ, each with a different cost model.
Analytical expressions of secrecy transmission capacity (STC) were also derived to depict the network security performance under the OWJ attack.
The results showed that the OWJ attack model causes more significant network security performance degradation than the pure wiretapping or jamming attack.
In addition, POWJ is the most hazardous, while DOWJ and LOWJ achieve almost the same attack effect.
This reveals that the effective antenna gain can be exploited to significantly improve the attack effect, whereas the link status has little impact on the improvement.

\appendices
\section{Proof of Lemma \ref{lemma:zeta-sad}}\label{app:proof-lemma-zeta-sad}  
Suppose $d_{x_0,z}=u$. 
We define $\tilde{\mathsf{D}}_W^z=\min_{x\in\Phi_T\backslash\{x_0\}}d_{x,z}$ and $\tilde{\mathsf{D}}_J^z=\min_{y\in\Phi_R\backslash\{y_0\}}d_{y,z}$.
The event $\mathbf{1}_J^z=1$ occurs in the following cases:
\begin{itemize}
\item $x_0$ is the nearest transmitter to $z$ (i.e., $\tilde{\mathsf{D}}_W^z\geq u$), $y_0$ is the nearest receiver to $z$ (i.e., $\tilde{\mathsf{D}}_J^z\geq v$) and $u\geq \rho v$;

\item $x_0$ is the nearest transmitter to $z$ (i.e., $\tilde{\mathsf{D}}_W^z\geq u$), $y_0$ is \emph{not} the nearest receiver  to $z$ (i.e., $\tilde{\mathsf{D}}_J^z<v$) and $u\geq \rho\tilde{\mathsf{D}}_J^z$;

\item $x_0$ is \emph{not} the nearest transmitter to $z$ (i.e., $\tilde{\mathsf{D}}_W^z< u$), $y_0$ is the nearest receiver to $z$ (i.e., $\tilde{\mathsf{D}}_J^z\geq v$) and $\tilde{\mathsf{D}}_W^z\geq \rho v$;

\item $x_0$ is \emph{not} the nearest transmitter to $z$ (i.e., $\tilde{\mathsf{D}}_W^z< u$), $y_0$ is \emph{not} the nearest receiver to $z$  (i.e., $\tilde{\mathsf{D}}_J^z<v$)  and $\tilde{\mathsf{D}}_W^z\geq \rho \tilde{\mathsf{D}}_J^z$.
\end{itemize}
Combining the four cases, we have $\mathbf{1}_J^z=1$  if
\begin{IEEEeqnarray}{rCl}\label{eqn:condofjamming-sad}
\begin{dcases}
\tilde{\mathsf{D}}_W^z\!\geq\! \rho \tilde{\mathsf{D}}_J^z,\! \tilde{\mathsf{D}}_J^z\!<\!\frac{u}{\rho}, &u<\rho v,\\
\tilde{\mathsf{D}}_W^z\!\geq\! \rho v, \!\tilde{\mathsf{D}}_J^z\!\geq\! v ~\mathrm{or}~ \tilde{\mathsf{D}}_W^z\!\geq\! \rho \tilde{\mathsf{D}}_J^z,\! \tilde{\mathsf{D}}_J^z\!<\!v, &u\geq \rho v.
\end{dcases}
\end{IEEEeqnarray}

Note that $\tilde{\mathsf{D}}_W^z$ (resp. $\tilde{\mathsf{D}}_J^z$) has the same PDF and CDF as those of $\mathsf{D}_W^z$ (resp. $\mathsf{D}_J^z$).
With the help of the PDFs and CCDFs of $\tilde{\mathsf{D}}_W^z$ and $\tilde{\mathsf{D}}_J^z$, we obtain the probability of $\mathbf{1}_J^z=1$ conditioned on $d_{x_0,z}=u$ as follows:
\begin{IEEEeqnarray}{rCl}\label{eqn:zeta_uv}
\zeta(u,v)
&=&\begin{dcases}
     \frac{1}{\rho^2+1}(1-e^{-(1 +\frac{1}{\rho^2})\lambda\pi u^2}), & u< \rho v\\
    \frac{1}{\rho^2+1}+\frac{\rho^2}{\rho^2+1}e^{-(\rho^2 +1)\lambda\pi v^2}, & u\ge \rho v
  \end{dcases}.
\end{IEEEeqnarray}
From \cite{YZhang2022TIFS}, we know that the CCDF and PDF of $d_{x_0,z}$ can be given by \eqref{eqn:ccdf_dx0z} and \eqref{eqn:pdf_dx0z}, respectively.
Calculating the expectation of \eqref{eqn:zeta_uv} in terms of $d_{x_0,z}$ yields $\zeta(v)$ in \eqref{eqn:zeta}.

\section{Proof of Lemma \ref{lemma:zeta-sal}}\label{app:proof-lemma-zeta-sal}  
Similar to the proof of Lemma \ref{lemma:zeta-sad} in Appendix \ref{app:proof-lemma-zeta-sad}, we first derive the probability of $\mathbf{1}_J^z=1$ (denoted by $\zeta_{\kappa}^{\tau}(u,v)$) given the distance $d_{x_0,z}$ and status $\mathcal S_{x_0,z}$ of the link $x_0\rightarrow z$.
Suppose $d_{x_0,z}=u$ and $\mathcal S_{x_0,z} = \kappa \in\{L,N\}$. 
It follows from Appendix \ref{app:proof-lemma-zeta-sad} that $\mathbf{1}_J^z=1$ if
\begin{IEEEeqnarray}{rCl}\label{eqn:condofjamming-sal}
\begin{dcases}
\tilde{\mathsf{L}}_T^z\!\geq\! \rho \tilde{\mathsf{L}}_R^z, \tilde{\mathsf{L}}_R^z \!<\!\frac{u^{\alpha_{\kappa}}}{\rho}, &u^{\alpha_{\kappa}}\!<\!\rho v^{\alpha_{\tau}},\\\\
\tilde{\mathsf{L}}_T^z \!\geq\! \rho v^{\alpha_{\tau}}, \!\tilde{\mathsf{L}}_R^z \!\geq\! v^{\alpha_{\tau}}~\mathrm{or}~\tilde{\mathsf{L}}_T^z \!\geq\! \rho \tilde{\mathsf{L}}_R^z, \tilde{\mathsf{L}}_R^z\!<\!v^{\alpha_{\tau}}, &u^{\alpha_{\kappa}}\!\geq\! \rho v^{\alpha_{\tau}},
\end{dcases}
\end{IEEEeqnarray}
where $\tilde{\mathsf{L}}_T^z=\min_{x\in\Phi_T\backslash\{x_0\}}d_{y,z}^{\alpha}$ and $\tilde{\mathsf{L}}_R^z=\min_{x\in\Phi_R\backslash\{y_0\}}d_{y,z}^{\alpha}$.
Next, we need to derive the PDFs and CCDFs of $\tilde{\mathsf{L}}_T^z$ and $\tilde{\mathsf{L}}_R^z$.
The CCDF of $\tilde{\mathsf{L}}_T^z$ is 
\begin{IEEEeqnarray}{rCl}\label{eqn:ccdfltz-derivation}
\mathbb P(\tilde{\mathsf{L}}_T^z>w)
&=&\mathbb P\left(\min_{x\in\Phi_T\backslash\{x_0\}} d_{x,z}^{\alpha}>w\right)\\
&=&\mathbb P\left(\bigcap_{x\in\Phi_T\backslash\{x_0\}} \{d_{x,z}^{\alpha}>w\}\right)\nonumber\\
&=&\mathbb E_{\Phi_T}\left[\prod_{x\in\Phi_T\backslash\{x_0\}}\mathbb P\left(d_{x,z}^{\alpha}>w\right)\right]\nonumber\\
&=&\exp\bigg(-2\pi\lambda\int_0^{\infty}(1-\mathbb P\left(r^{\alpha}>w\right))r\mathrm d r \bigg)\nonumber\\
&\overset{(b)}{=}&e^{-\Lambda(\lambda, w)}\nonumber,
\end{IEEEeqnarray}
where $(b)$ follows after chaning $\mathbb P\left(r^{\alpha}>w\right)$ to $p_L(r)\mathbf{1}_{r^{\alpha_L}>w}+p_N(r)\mathbf{1}_{r^{\alpha_N}>w}$.
Note that the CCDF of $\tilde{\mathsf{L}}_R^z$ is identical to that of  $\tilde{\mathsf{L}}_T^z$. 
Hence, the PDF of $\tilde{\mathsf{L}}_R^z$ is
\begin{IEEEeqnarray}{rCl}\label{eqn:pdflrz-derivation}
 f_{\tilde{\mathsf{L}}_R^z}( w )=e^{-\Lambda(\lambda, w)}\Lambda'(\lambda , w ).
\end{IEEEeqnarray}
Thus, for $u^{\alpha_{\kappa}}<\rho v^{\alpha_{\tau}}$, the probability of $\mathbf{1}_J^z=1$ is
\begin{IEEEeqnarray}{rCl}
\zeta_{\kappa}^{\tau}(u,v)&=&\int_{0}^{\frac{u^{\alpha_{\kappa}}}{\rho}}\mathbb P(\tilde{\mathsf{L}}_T^z\geq \rho  w ) f_{\tilde{\mathsf{L}}_R^z}( w )\mathrm d  w \\
&=&\int_{0}^{\frac{u^{\alpha_{\kappa}}}{\rho}}e^{-(\Lambda(\lambda, \rho  w )+\Lambda(\lambda, w ))} \Lambda'(\lambda , w )\mathrm d  w \nonumber.
\end{IEEEeqnarray}
For $u^{\alpha_{\kappa}}\geq \rho v^{\alpha_{\tau}}$, the probability is
\begin{IEEEeqnarray}{rCl}
\zeta_{\kappa}^{\tau}(u,v)&=&\mathbb P(\tilde{\mathsf{L}}_T^z\geq \rho v^{\alpha_{\tau}}, \tilde{\mathsf{L}}_R^z\geq v^{\alpha_{\tau}})\\
&&+\mathbb P(\tilde{\mathsf{L}}_T^z\geq \rho \tilde{\mathsf{L}}_R^z, \tilde{\mathsf{L}}_R^z<v^{\alpha_{\tau}})\nonumber\\
&=&e^{-(\Lambda(\lambda, \rho v^{\alpha_{\tau}})+\Lambda(\lambda, v^{\alpha_{\tau}}))}\nonumber\\
&&+\int_0^{v^{\alpha_{\tau}}} e^{-(\Lambda(\lambda, \rho  w)+\Lambda(\lambda, w))} \Lambda'(\lambda, w)\mathrm d  w \nonumber\\
&=&1-\rho \int_0^{v^{\alpha_{\tau}}} e^{-(\Lambda(\lambda, \rho  w)+\Lambda(\lambda,  w))} \Lambda'(\lambda, \rho w)\mathrm d  w \nonumber.
\end{IEEEeqnarray}

Finally, taking the expectation of  $\zeta_{\kappa}^{\tau}(u,v)$ in terms of $d_{x_0,z}$ and $\mathcal S_{x_0,z}$ completes the proof.
%

\section{Proof of Lemma \ref{lemma:zeta-sap}}\label{app:proof-lemma-zeta-sap}  
Following the idea in Appendix \ref{app:proof-lemma-zeta-sal}, we first derive the probability of $\mathbf{1}_J^z=1$ (denoted by $\zeta_{\kappa,l,m}^{\tau,n,o}(u,v)$) conditioned on $d_{x_0,z}=u$, status $\mathcal S_{x_0,z}=\kappa$, effective antenna gain $\mathsf{G}_{T,W}^{x_0, z} = G_l^TG_m^W$ of the link $x_0\rightarrow z$. 
From Appendix \ref{app:proof-lemma-zeta-sad}, we can see that $\mathbf{1}_J^z=1$ if 
\begin{IEEEeqnarray}{rCl}\label{eqn:condofjamming-sap}
\begin{dcases}
\tilde{\mathsf{P}}_T^z\!\geq\! \rho \tilde{\mathsf{P}}_R^z, \tilde{\mathsf{P}}_R^z\!<\!\frac{\eta_1}{\rho}, &\!\!\eta_1\!<\! \rho\eta_2,\\
\tilde{\mathsf{P}}_T^z\!\geq\!  \rho\eta_2, \tilde{\mathsf{P}}_R^z\!\geq\!\eta_2~\mathrm{or}~ \tilde{\mathsf{P}}_T^z\!\geq\! \rho \tilde{\mathsf{P}}_R^z, \tilde{\mathsf{P}}_R^z\!<\!\eta_2, &\!\!\eta_1\!\ge\! \rho\eta_2,
\end{dcases}
\end{IEEEeqnarray}
where $\tilde{\mathsf{P}}_T^z=\min_{x\in\Phi_T\backslash\{x_0\}}d_{x,z}^{\alpha}/(P_T \mathsf{G}_{T,W})$ and $\tilde{\mathsf{L}}_R^z=\min_{x\in\Phi_R\backslash\{y_0\}}d_{y,z}^{\alpha}/(P_J \mathsf{G}_{J,R})$.

Letting $\tilde{\mathsf{L}}_{T}^{z,i,j}=\min_{x\in\Phi_T^{i,j}\backslash\{x_0\}}d_{x,z}^{\alpha}$, where $\Phi_T^{i,j}$ is the PPP of transmitters with  antenna gain $G_i^T G_j^W$ to $z$, we have
\begin{IEEEeqnarray}{rCl}\label{eqn:ccdf-ptz}
\mathbb P(\tilde{\mathsf{P}}_T^z>w)&=&\prod_{i,j}\mathbb P\left(\tilde{\mathsf{L}}_{T}^{z,i,j}>wP_TG_i^TG_j^W\right)\\
&\overset{(c)}{=}&\prod_{i,j}e^{-\Lambda(p_{ij}^{TW}\lambda, P_TG_i^T G_j^W w)}\nonumber\\
&=&\exp\left(-\sum_{i,j}\Lambda(p_{ij}^{TW}\lambda, P_TG_i^T G_j^W w)\right)\nonumber\\
&=&e^{-\hat{\Lambda}_1(\lambda, w)}\nonumber,
\end{IEEEeqnarray}
where $i\in\{M,S\},j\in\{M,S\}$ and (c) follows from \eqref{eqn:ccdfltz-derivation}.
Similarly, we have
\begin{IEEEeqnarray}{rCl}\label{eqn:ccdf-prz}
\mathbb P(\tilde{\mathsf{P}}_R^z>w)&=&e^{-\hat{\Lambda}_2(\lambda, w)}.
\end{IEEEeqnarray}
The PDFs of $\tilde{\mathsf{P}}_T^z$ and $\tilde{\mathsf{P}}_R^z$ can de derived accordingly.
Based on the PDFs and CCDFs of $\tilde{\mathsf{P}}_T^z$ and $\tilde{\mathsf{P}}_R^z$, we can obtain the probability $\zeta_{\kappa,l,m}^{\tau,n,o}(u,v)$  as in \eqref{eqn:zetakappalmtauno1} and \eqref{eqn:zetakappalmtauno2}.
Finally, neglecting the dependence between $\mathsf{G}_{T,W}^{x_0,z}$ and $\mathsf{G}_{J,R}^{z,y_0}$ and taking the expectation of $\zeta_{\kappa,l,m}^{\tau,n,o}(u,v)$ in terms of $d_{x_0,z}$, $\mathcal S_{x_0,z}$ and $\mathsf{G}_{T,W}^{x_0, z}$, we complete the proof.

\section{Proof of Lemma \ref{lemma:ltzt-sad}}\label{app:lemma-ltzt-sad}
We first calculate the Laplace transform of $\hat{I}_z^T$ conditioned  on $d_{y_0,z}=v$, which is given by 
\begin{IEEEeqnarray}{rCl}\label{eqn:ltztsuv}
\mathcal L_z^T(s,u,v)&=&\mathbb E[e^{-s\hat{I}_z^T}]\\
&=&\mathbb E[e^{-s\hat{I}_z^T} | \mathbf{1}_J^z=1]\mathbb P(\mathbf{1}_J^z=1)\nonumber\\
&&+\mathbb E[e^{-s\hat{I}_z^T} | \mathbf{1}_J^z=0]\mathbb P(\mathbf{1}_J^z=0)\nonumber\\
&=&\mathbb E[e^{-sI_z^T} | \mathbf{1}_J^z=0]\mathbb P(\mathbf{1}_J^z=0)\nonumber\\
&=&\mathbb E[e^{-sI_z^T}]-\mathbb E[e^{-sI_z^T} | \mathbf{1}_J^z=1]\mathbb P(\mathbf{1}_J^z=1)\nonumber.
\end{IEEEeqnarray}
Following from Lemma \ref{lemma:ly0t}, $\mathbb E[e^{-sI_z^T}]$ can be given by
\begin{IEEEeqnarray}{rCl}
\mathbb E[e^{-sI_z^T}]=\Xi_{T,W}[s,\lambda,0].
\end{IEEEeqnarray}
According to the conditions of $\mathbf{1}_J^z=1$ in \eqref{eqn:condofjamming-sad}, we have 
\begin{IEEEeqnarray}{rCl} \label{eqn:ltztsuv-proof1}
\mathcal L_z^T(s,u,v)&=&\mathbb E[e^{-sI_z^T}]\\
&&\!\!-\!\!\int_{0}^{\frac{u}{\rho}}\!\!\!\!\mathbb E[e^{-sI_z^T}|\tilde{\mathsf{D}}_T^z\geq \rho  w ]\mathbb P(\tilde{\mathsf{D}}_T^z\geq \rho  w )f_{\tilde{\mathsf{D}}_R^z}( w )\mathrm d  w \nonumber\\
&=&\Xi_{T,W}(s,\lambda,0)\nonumber\\
&&-\int_{0}^{\frac{u}{\rho}}\Xi_{T,W}(s,\lambda,\rho  w )2\pi\lambda  w e^{-(\rho^2+1)\lambda\pi w ^2} \mathrm d   w\nonumber
\end{IEEEeqnarray}
for $u<\rho v$, and
\begin{IEEEeqnarray}{rCl} \label{eqn:ltztsuv-proof2}
\mathcal L_z^T(s,u,v)&=&\mathbb E[e^{-sI_z^T}]\\
&&-\mathbb E[e^{-sI_z^T}|\tilde{\mathsf{D}}_T^z\!\ge \!\rho v]\mathbb P(\tilde{\mathsf{D}}_T^z\!\ge\! \rho v) \mathbb P(\tilde{\mathsf{D}}_R^z\!\geq\! v)\nonumber\\
&&-\!\!\int_{0}^{v}\!\!\mathbb E[e^{-sI_z^T}|\tilde{\mathsf{D}}_T^z\!\geq\!\rho w ]\mathbb P(\tilde{\mathsf{D}}_T^z\!\geq\! \rho  w )f_{\tilde{\mathsf{D}}_R^z}( w )\mathrm d  w \nonumber\\
&=&\Xi_{T,W}(s,\lambda,0)\!-\!e^{-(\rho^2+1)\lambda \pi v^2}\Xi_{T,W}(s,\lambda,\rho v)\nonumber\\
&&-\!\!\int_{0}^{v}\Xi_{T,W}(s,\lambda,\rho  w )2\pi\lambda  w e^{-(\rho^2+1)\lambda\pi w ^2} \mathrm d   w \nonumber
\end{IEEEeqnarray}
for  $u\geq \rho v$.
Taking the expectation of $\mathcal L_z^T(s,u,v)$ in terms of $d_{y_0,z}$ completes the proof.

\section{Proof of Lemma \ref{lemma:ltzt-sal}}\label{app:lemma-ltzt-sal}
We first calculate the Laplace transform conditioned on $d_{y_0,z}=v$ and $\mathcal S_{y_0,z}=\tau$ (denoted by $\mathcal L_{z,\kappa}^{T,\tau}(s,u,v)$).
According to \eqref{eqn:ltztsuv}, we need to derive $\mathbb E[e^{-sI_z^T} | \mathbf{1}_J^z=1]\mathbb P(\mathbf{1}_J^z=1)$.
We rewrite $I_z^T$ as
\begin{IEEEeqnarray}{rCl}
I_z^T=\sum_{k\in\{L,N\}}I_z^{T,k},
\end{IEEEeqnarray}
where $I_z^{T,k}=\sum_{x\in\Phi_T^k\backslash\{x_0\}} P_T\mathsf{G}_{T,W}^{x,z} h_{x,z}d_{x,z}^{-\alpha}$ denotes the interference from the sub-PPP $\Phi_T^k$ of transmitters with link status $k$ to $z$.

Defining $\tilde{\mathsf{D}}_{T,k}^z=\min_{x\in \Phi_T^k\backslash\{x_0\}} d_{x,z}$, we have
\begin{IEEEeqnarray}{rCl}
\mathcal L_{z,\kappa}^{T,\tau}(s,u,v)&=&\mathbb E[e^{-sI_z^T}]-\int_{0}^{\frac{u^{\alpha_{\kappa}}}{\rho}}\!\!\!\mathbb P(\tilde{\mathsf{L}}_T^z\geq \rho w) f_{\tilde{\mathsf{L}}_R^z}( w )\\
&&\quad \quad \quad \prod_{k}\mathbb E[e^{-sI_z^{T,k}}|\tilde{\mathsf{L}}_T^z\geq \rho w]\mathrm d  w\nonumber\\
&=&\mathbb E[e^{-sI_z^T}]-\int_{0}^{\frac{u^{\alpha_{\kappa}}}{\rho}}\mathbb P(\tilde{\mathsf{L}}_T^z\geq \rho w) f_{\tilde{\mathsf{L}}_R^z}( w )\nonumber\\
&&\quad \prod_{k}\mathbb E[e^{-sI_z^{T,k}}|\tilde{\mathsf{D}}_{T,k}^z\geq (\rho  w)^{\frac{1}{\alpha_k}}]\mathrm d  w\nonumber
\end{IEEEeqnarray}
for $u^{\alpha_{\kappa}}<\rho v^{\alpha_{\tau}}$, and
\begin{IEEEeqnarray}{rCl}
\mathcal L_{z,\kappa}^{T,\tau}(s,u,v)&=&\mathbb E[e^{-sI_z^T}]-\mathbb E[e^{-sI_z^T}|\tilde{\mathsf{L}}_T^z\geq \rho v^{\alpha_{\tau}}]\\
&&\quad \quad \quad \times\mathbb P(\tilde{\mathsf{L}}_T^z\geq \rho v^{\alpha_{\tau}}, \tilde{\mathsf{L}}_R^z\geq v^{\alpha_{\tau}})\nonumber\\
&&-\int_0^{v^{\alpha_{\tau}}}\mathbb P(\tilde{\mathsf{L}}_T^z\geq \rho w)f_{\tilde{\mathsf{L}}_R^z}( w ) \nonumber\\
&&\quad \quad \quad \prod_{k}\mathbb E[e^{-sI_z^{T,k}}|\tilde{\mathsf{D}}_{T,k}^z\geq (\rho  w)^{\frac{1}{\alpha_k}}]\mathrm dw \nonumber
\end{IEEEeqnarray}
for $u^{\alpha_{\kappa}}\geq \rho v^{\alpha_{\tau}}$. After some mathematical manipulations based on \eqref{eqn:ltztsuv-proof1} and \eqref{eqn:ltztsuv-proof2}, we obtain the expression of $\mathcal L_{z,\kappa}^{T,\tau}(s,u,v)$ in \eqref{eqn:ltztkappatau}. 
We complete the proof after calculating the expectation of $\mathcal L_{z,\kappa}^{T,\tau}(s,u,v)$  in terms of $d_{y_0,z}$ and $\mathcal S_{y_0,z}$.

\section{Proof of Lemma \ref{lemma:ltzt-sap}}\label{app:lemma-ltzt-sap}
Similar to Appendix \ref{app:lemma-ltzt-sal}, we first derive the Laplace transform conditioned on $d_{y_0,z}=v$, $\mathcal S_{y_0,z}=\tau$ and $\mathsf{G}_{J,R}^{z,y_0}=G_n^JG_o^R$ (denoted by $\mathcal L_{z,\kappa,l,m}^{T,\tau,n,o}(s,u,v)$).
We rewrite $I_z^T$ as
\begin{IEEEeqnarray}{rCl}
I_z^T=\sum_{k\in\{L,N\}}\sum_{i\in\{M,S\}}\sum_{j\in\{M,S\}}I_z^{T,k,i,j},
\end{IEEEeqnarray}
where $I_z^{T,k,i,j}=\sum_{x\in\Phi_T^{k,i,j}\backslash\{x_0\}} P_T\mathsf{G}_{T,W}^{x,z} h_{x,z}d_{x,z}^{-\alpha}$ denotes the interference from the sub-PPP $\Phi_T^{k,i,j}$ of transmitters with link status $k$ and channel gain $G_i^TG_j^W$to $z$.

Let $\tilde{\mathsf{D}}_{T,k,i,j}^z=\min_{x\in \tilde{\mathsf{D}}_{T,k,i,j}^z\backslash\{x_0\}} d_{x,z}$.
According to \eqref{eqn:condofjamming-sap}, for $\eta_1<\rho\eta_2$, the Laplace transform $\mathcal L_{z,\kappa,l,m}^{T,\tau,n,o}(s,u,v)$ is
\begin{IEEEeqnarray}{rCl}
&&\mathbb E[e^{-sI_z^T}]-\int_{0}^{\frac{\eta_1}{\rho}}\mathbb P(\tilde{\mathsf{P}}_T^z\geq \rho w)f_{\tilde{\mathsf{P}}_R^z}(w)\\
&& \times \prod_{k,i,j}\mathbb E[e^{-sI_z^{T,k,i,j}}|\tilde{\mathsf{D}}_{T,k,i,j}^z\geq (\rho  w P_T G_i^T G_j^W)^{\frac{1}{\alpha_k}}]\mathrm d  w\nonumber.
\end{IEEEeqnarray}

For  $\eta_1\ge \rho\eta_2$, the Laplace transform $\mathcal L_{z,\kappa,l,m}^{T,\tau,n,o}(s,u,v)$ is 
\begin{IEEEeqnarray}{rCl}
&&\mathbb E[e^{-sI_z^T}]-\mathbb E\left[e^{-sI_z^T}|\tilde{\mathsf{P}}_T^z\geq \rho\eta_2\right] \mathbb P\left(\tilde{\mathsf{P}}_T^z\geq  \rho\eta_2, \tilde{\mathsf{P}}_R^z\geq\eta_2\right)\nonumber\\
&&-\int_0^{\eta_2}\mathbb P(\tilde{\mathsf{P}}_T^z\geq \rho w)f_{\tilde{\mathsf{P}}_R^z}(w)\nonumber\\
&& \times \prod_{k,i,j}\mathbb E[e^{-sI_z^{T,k,i,j}}|\tilde{\mathsf{D}}_{T,k,i,j}^z\geq (\rho  w P_T G_i^T G_j^W)^{\frac{1}{\alpha_k}}]\mathrm d  w.
\end{IEEEeqnarray}
We then obtain the expression of $\mathcal L_{z,\kappa,l,m}^{T,\tau,n,o}(s,u,v)$ in \eqref{eqn:ltztkappalmtauno} after conducting some mathematical manipulations.
Calculating the expectation of $\mathcal L_{z,\kappa,l,m}^{T,\tau,n,o}(s,u,v)$ in terms of $d_{y_0,z}$, $\mathcal S_{y_0,z}$ and $\mathsf{G}_{J,R}^{z,y_0}$ completes the proof.

\bibliographystyle{IEEEtran}
\bibliography{references.bib}

\begin{thebibliography}{10}
\providecommand{\url}[1]{#1}
\csname url@samestyle\endcsname
\providecommand{\newblock}{\relax}
\providecommand{\bibinfo}[2]{#2}
\providecommand{\BIBentrySTDinterwordspacing}{\spaceskip=0pt\relax}
\providecommand{\BIBentryALTinterwordstretchfactor}{4}
\providecommand{\BIBentryALTinterwordspacing}{\spaceskip=\fontdimen2\font plus
\BIBentryALTinterwordstretchfactor\fontdimen3\font minus
  \fontdimen4\font\relax}
\providecommand{\BIBforeignlanguage}[2]{{%
\expandafter\ifx\csname l@#1\endcsname\relax
\typeout{** WARNING: IEEEtran.bst: No hyphenation pattern has been}%
\typeout{** loaded for the language `#1'. Using the pattern for}%
\typeout{** the default language instead.}%
\else
\language=\csname l@#1\endcsname
\fi
#2}}
\providecommand{\BIBdecl}{\relax}
\BIBdecl

\bibitem{Rappaport2013IEEEAccess}
T.~S. {Rappaport}, S.~{Sun}, R.~{Mayzus}, H.~{Zhao}, Y.~{Azar}, K.~{Wang},
  G.~N. {Wong}, J.~K. {Schulz}, M.~{Samimi}, and F.~{Gutierrez}, ``Millimeter
  wave mobile communications for 5g cellular: It will work!'' \emph{IEEE
  Access}, vol.~1, pp. 335--349, 2013.

\bibitem{RanganS2014IEEEProceedings}
S.~Rangan, T.~S. Rappaport, and E.~Erkip, ``Millimeter-wave cellular wireless
  networks: Potentials and challenges,'' \emph{Proceedings of the IEEE}, vol.
  102, no.~3, pp. 366--385, 2014.

\bibitem{Hong2021JMicro}
W.~Hong, Z.~H. Jiang, C.~Yu, D.~Hou, H.~Wang, C.~Guo, Y.~Hu, L.~Kuai, Y.~Yu,
  Z.~Jiang, Z.~Chen, J.~Chen, Z.~Yu, J.~Zhai, N.~Zhang, L.~Tian, F.~Wu,
  G.~Yang, Z.-C. Hao, and J.~Y. Zhou, ``The role of millimeter-wave
  technologies in 5g/6g wireless communications,'' \emph{IEEE J. Microwaves},
  vol.~1, no.~1, pp. 101--122, 2021.

\bibitem{ZXiao2022IEEECST}
Z.~Xiao, L.~Zhu, Y.~Liu, P.~Yi, R.~Zhang, X.-G. Xia, and R.~Schober, ``A survey
  on millimeter-wave beamforming enabled uav communications and networking,''
  \emph{{IEEE} Commun. Surveys Tuts.}, vol.~24, no.~1, pp. 557--610, 2022.

\bibitem{ZGuo2021IEEETAP}
Z.-J. Guo, Z.-C. Hao, H.-Y. Yin, D.-M. Sun, and G.~Q. Luo, ``Planar
  shared-aperture array antenna with a high isolation for millimeter-wave low
  earth orbit satellite communication system,'' \emph{IEEE Trans. Antennas
  Propag.}, vol.~69, no.~11, pp. 7582--7592, 2021.

\bibitem{Zhang2019TCOM}
Y.~{Zhang}, Y.~{Shen}, X.~{Jiang}, and S.~{Kasahara}, ``Mode selection and
  spectrum partition for d2d inband communications: A physical layer security
  perspective,'' \emph{IEEE Trans. Commun.}, vol.~67, no.~1, pp. 623--638, Jan
  2019.

\bibitem{SZhao2021TIFS}
S.~Zhao, J.~Liu, Y.~Shen, X.~Jiang, and N.~Shiratori, ``Secure and
  energy-efficient precoding for mimo two-way untrusted relay systems,''
  \emph{{IEEE} Trans. Inf. Forensics Security}, vol.~16, pp. 3371--3386, 2021.

\bibitem{YXu2021TWC}
Y.~Xu, J.~Liu, Y.~Shen, X.~Jiang, Y.~Ji, and N.~Shiratori, ``Qos-aware secure
  routing design for wireless networks with selfish jammers,'' \emph{IEEE
  Transactions on Wireless Communications}, vol.~20, no.~8, pp. 4902--4916,
  2021.

\bibitem{SZhao2020TIFS}
S.~Zhao, J.~Liu, Y.~Shen, X.~Jiang, and N.~Shiratori, ``Secure beamforming for
  full-duplex mimo two-way untrusted relay systems,'' \emph{{IEEE} Trans. Inf.
  Forensics Security}, vol.~15, pp. 3775--3790, 2020.

\bibitem{JHe2020TIFS}
J.~He, J.~Liu, Y.~Shen, X.~Jiang, and N.~Shiratori, ``Link selection for
  security-qos tradeoffs in buffer-aided relaying networks,'' \emph{{IEEE}
  Trans. Inf. Forensics Security}, vol.~15, pp. 1347--1362, 2020.

\bibitem{ZKong2021TIFS}
Z.~Kong, J.~Song, C.~Wang, H.~Chen, and L.~Hanzo, ``Hybrid analog-digital
  precoder design for securing cognitive millimeter wave networks,''
  \emph{{IEEE} Trans. Inf. Forensics Security}, vol.~16, pp. 4019--4034, 2021.

\bibitem{JXU2019TCOM}
J.~Xu, W.~Xu, D.~W.~K. Ng, and A.~L. Swindlehurst, ``Secure communication for
  spatially sparse millimeter-wave massive mimo channels via hybrid
  precoding,'' \emph{IEEE Trans. Commun.}, vol.~68, no.~2, pp. 887--901, 2020.

\bibitem{KWu2020ComLett}
K.~Wu, W.~Ni, J.~A. Zhang, R.~P. Liu, and J.~Guo, ``Secrecy rate analysis for
  millimeter-wave lens antenna array transmission,'' \emph{{IEEE} Commun.
  Lett.}, vol.~24, no.~2, pp. 272--276, 2020.

\bibitem{WHao2020TCOM}
W.~Hao, G.~Sun, J.~Zhang, P.~Xiao, and L.~Hanzo, ``Secure millimeter wave cloud
  radio access networks relying on microwave multicast fronthaul,'' \emph{IEEE
  Trans. Commun.}, vol.~68, no.~5, pp. 3079--3095, 2020.

\bibitem{JSong2020TvT}
J.~Song, B.~Lee, J.~Park, M.-S. Lee, and J.-H. Lee, ``Beamformer design for
  physical layer security in dual-polarized millimeter wave channels,''
  \emph{IEEE Trans. Veh. Technol.}, vol.~69, no.~10, pp. 12\,306--12\,311,
  2020.

\bibitem{SHuang2020TvT}
S.~Huang, M.~Xiao, and H.~V. Poor, ``On the physical layer security of
  millimeter wave noma networks,'' \emph{IEEE Trans. Veh. Technol.}, vol.~69,
  no.~10, pp. 11\,697--11\,711, 2020.

\bibitem{YSong2020TvT}
Y.~Song, W.~Yang, Z.~Xiang, H.~Wang, and F.~Cao, ``Research on cognitive power
  allocation for secure millimeter-wave noma networks,'' \emph{IEEE Trans. Veh.
  Technol.}, vol.~69, no.~11, pp. 13\,424--13\,436, 2020.

\bibitem{YYapici2021TvT}
Y.~Yapici, N.~Rupasinghe, I.~Guvenc, H.~Dai, and A.~Bhuyan, ``Physical layer
  security for noma transmission in mmwave drone networks,'' \emph{IEEE Trans.
  Veh. Technol.}, vol.~70, no.~4, pp. 3568--3582, 2021.

\bibitem{XSun2020IoTJ}
X.~Sun, W.~Yang, and Y.~Cai, ``Secure communication in noma-assisted
  millimeter-wave swipt uav networks,'' \emph{IEEE Internet of Things Journal},
  vol.~7, no.~3, pp. 1884--1897, 2020.

\bibitem{YZhu2020JSAC}
Y.~Zhu, G.~Zheng, and K.-K. Wong, ``Stochastic geometry analysis of large
  intelligent surface-assisted millimeter wave networks,'' \emph{IEEE J. Sel.
  Areas Commun.}, vol.~38, no.~8, pp. 1749--1762, 2020.

\bibitem{CWang2021TvT}
C.~Wang, Z.~Li, T.-X. Zheng, D.~W.~K. Ng, and N.~Al-Dhahir, ``Intelligent
  reflecting surface-aided secure broadcasting in millimeter wave symbiotic
  radio networks,'' \emph{IEEE Trans. Veh. Technol.}, vol.~70, no.~10, pp.
  11\,050--11\,055, 2021.

\bibitem{YXiu2021ComLett}
Y.~Xiu, J.~Zhao, W.~Sun, and Z.~Zhang, ``Secrecy rate maximization for
  reconfigurable intelligent surface aided millimeter wave system with
  low-resolution dacs,'' \emph{{IEEE} Commun. Lett.}, vol.~25, no.~7, pp.
  2166--2170, 2021.

\bibitem{Shakir2022PhotonicsJ}
W.~M.~R. Shakir and M.-S. Alouini, ``Secrecy performance analysis of parallel
  fso/mm-wave system over unified fisher-snedecor channels,'' \emph{IEEE
  Photonics Journal}, vol.~14, no.~2, pp. 1--13, 2022.

\bibitem{Tokgoz2022TvT}
S.~C. Tokgoz, S.~Althunibat, S.~L. Miller, and K.~A. Qaraqe, ``On the secrecy
  capacity of hybrid fso-mmwave wiretap channels,'' \emph{IEEE Trans. Veh.
  Technol.}, vol.~71, no.~4, pp. 4073--4086, 2022.

\bibitem{Tokgoz2021ICC}
S.~C. Tokgoz, S.~Althunibat, S.~Yarkan, and K.~A. Qaraqe, ``Physical layer
  security of hybrid fso-mmwave communications in presence of correlated
  wiretap channels,'' in \emph{Proc. IEEE ICC}, 2021, pp. 1--7.

\bibitem{Saber2021SystJ}
M.~J. Saber and S.~Rajabi, ``On secrecy performance of millimeter-wave
  rf-assisted fso communication systems,'' \emph{IEEE Syst. J.}, vol.~15,
  no.~3, pp. 3781--3788, 2021.

\bibitem{YZhang2022TIFS}
Y.~Zhang, Y.~Shen, X.~Jiang, and S.~Kasahara, ``Secure millimeter-wave ad hoc
  communications using physical layer security,'' \emph{{IEEE} Trans. Inf.
  Forensics Security}, vol.~17, pp. 99--114, 2022.

\bibitem{YZhu2017TWC}
Y.~{Zhu}, L.~{Wang}, K.~{Wong}, and R.~W. {Heath}, ``Secure communications in
  millimeter wave ad hoc networks,'' \emph{IEEE Trans. Wireless Commun.},
  vol.~16, no.~5, pp. 3205--3217, 2017.

\bibitem{YZhu2018JSAC}
Y.~{Zhu}, G.~{Zheng}, and M.~{Fitch}, ``Secrecy rate analysis of uav-enabled
  mmwave networks using matern hardcore point processes,'' \emph{IEEE J. Sel.
  Areas Commun.}, vol.~36, no.~7, pp. 1397--1409, 2018.

\bibitem{Rappaport2017TAP}
T.~S. {Rappaport}, Y.~{Xing}, G.~R. {MacCartney}, A.~F. {Molisch},
  E.~{Mellios}, and J.~{Zhang}, ``Overview of millimeter wave communications
  for fifth-generation (5g) wireless networks---with a focus on propagation
  models,'' \emph{IEEE Trans. Antennas Propag.}, vol.~65, no.~12, pp.
  6213--6230, 2017.

\bibitem{ma2018security}
J.~Ma, R.~Shrestha, J.~Adelberg, C.-Y. Yeh, Z.~Hossain, E.~Knightly, J.~M.
  Jornet, and D.~M. Mittleman, ``Security and eavesdropping in terahertz
  wireless links,'' \emph{Nature}, vol. 563, no. 7729, pp. 89--93, 2018.

\bibitem{akyildiz2014terahertz}
I.~F. Akyildiz, J.~M. Jornet, and C.~Han, ``Terahertz band: Next frontier for
  wireless communications,'' \emph{Physical communication}, vol.~12, pp.
  16--32, 2014.

\bibitem{harvey2019exploiting}
J.~F. Harvey, M.~B. Steer, and T.~S. Rappaport, ``Exploiting high millimeter
  wave bands for military communications, applications, and design,''
  \emph{IEEE Access}, vol.~7, pp. 52\,350--52\,359, 2019.

\bibitem{YZhang2021NaNA}
Y.~Zhang, J.~He, Q.~Qu, and Z.~Zhang, ``Wiretapping or jamming: On eavesdropper
  attacking strategy in mmwave ad hoc networks,'' in \emph{Proc. IEEE NaNA},
  2021, pp. 8--14.

\bibitem{haenggi2012stochastic}
M.~Haenggi, \emph{Stochastic geometry for wireless networks}.\hskip 1em plus
  0.5em minus 0.4em\relax Cambridge University Press, 2012.

\bibitem{AThornburg2016TSP}
A.~{Thornburg}, T.~{Bai}, and R.~W. {Heath}, ``Performance analysis of outdoor
  mmwave ad hoc networks,'' \emph{{IEEE} Trans. Signal Process.}, vol.~64,
  no.~15, pp. 4065--4079, 2016.

\end{thebibliography}


\end{document}